\documentclass[epj]{svjour}
\newcommand{\be}{\begin{equation}}
\newcommand{\ee}{\end{equation}}
\newcommand{\lb}[1]{\label{#1}}
\newcommand{\sty}{\scriptstyle}

\usepackage{epsfig}
\input{epsf}
\usepackage{txfonts}
\usepackage{graphics}
\begin{document}
\title{Evidence for the Gompertz Curve in the Income Distribution of
       Brazil 1978--2005}
\author{Newton J.\ Moura Jr\inst{1} \and Marcelo B.\ Ribeiro\inst{2}
       \thanks{Corresponding author}
       }                     
\institute{IBGE -- Brazilian Institute for Geography and Statistics,
           Geosciences Directorate, Geodesics Department, Av.\ Brasil
	   15671, Rio de Janeiro, RJ 21241-051, Brazil; e-mail:
	   newtonjunior@ibge.gov.br
	   \and
           Physics Institute, University of Brazil -- UFRJ, CxP 68532,
           Rio de Janeiro, RJ 21941-972, Brazil; e-mail: mbr@if.ufrj.br
	   }
\date{Received: / Revised version: }
\abstract{This work presents an empirical study of the evolution of
          the personal income distribution in Brazil. Yearly samples
	  available from 1978 to 2005 were studied and evidence was
	  found that the complementary cumulative distribution of
	  personal income for 99\% of the economically less favorable
	  population is well represented by a Gompertz curve of the
	  form $G\,(x)=\exp\,[\,\exp \, (A-Bx)]$, where $x$ is the
	  normalized individual income. The complementary cumulative
	  distribution of the remaining 1\% richest part of the
	  population is well represented by a Pareto power law
	  distribution $P\,(x)= \beta\,x^{-\alpha}$. This result
	  means that similarly to other countries, Brazil's income
	  distribution is characterized by a well defined two class
	  system. The parameters $A$, $B$, $\alpha$, $\beta$ were
	  determined by a mixture of boundary conditions, normalization
          and fitting methods for every year in the time span of
          this study. Since the Gompertz curve is characteristic of
	  growth models, its presence here suggests that these
	  patterns in income distribution could be a consequence of
	  the growth dynamics of the underlying economic system. In
	  addition, we found out that the percentage share of both
	  the Gompertzian and Paretian components relative to the
	  total income shows an approximate cycling pattern with
	  periods of about 4 years and whose maximum and minimum
	  peaks in each component alternate at about every 2 years.
	  This finding suggests that the growth dynamics of
	  Brazil's economic system might possibly follow a
	  Goodwin-type class 
	  model dynamics based on the application of the
	  Lotka-Volterra equation to economic growth and cycle.
} 
\maketitle

\section{Introduction}\label{intro}

The attempt to apply methods of physics to describe various features
of societies has a long history, stretching as far back as Thomas Hobbes
and William Petty \cite{b02,b04}. Although the potential for
misapplications is far from negligible \cite{b04,m89}, over the past
two decades tools, methods and ideas originally developed to understand
the fabric of the physical universe are being increasingly applied by
physicists to describe and understand the inner workings of societies
\cite{s05,ds08,doyne05,ga04,ms00,mc04,ri07,r02}. What started simply
as an exercise in statistical mechanics, where complex behavior arises
from simple rules caused by the interaction of a large number of
components, due to the increasing interest of physicists in
interdisciplinary research these applications have been constantly
growing and the area of what today is named as socio-economical
physics, sociophysics and econophysics for short, was born in the
late 1990s \cite{b06,cks07,ma07,nat06,y08}. As a
consequence, old problems in what until recently was believed to be the
exclusive realm of economics are receiving fresh attention in
econophysics and possible new perspectives and solutions are emerging.

Our goal here is to focus in one of those old problems, namely in the
work made over a century ago by the Italian economist and sociologist
Vilfredo Pareto \cite{pareto}, who studied the personal income
distribution for some countries and years. He found out that the
complementary cumulative personal income distributions followed a
power law for those with high income \cite[p.\ 245]{b04},
\cite[p.\ 152]{mh04}, a result that turned out later to be considered
a classic example of a fractal distribution \cite[p.\ 347]{mandelbrot},
\cite{n05}. Later results confirmed Pareto's findings, but the
application of his personal income power law, also known simply as
\textit{Pareto law} \cite{k80,n05}, is limited to the very high
income population (see below). The overwhelming majority of the
population does not follow Pareto's power law distribution and,
therefore, the characterization and understanding of the personal
income distribution of the economically less favored still remains an
open problem. 

There has been several recent studies about individual income distribution
for different countries and epochs, modern, medieval and even ancient.
For old societies, these studies include ancient Egypt \cite{a02} and
medieval Hungary around 1550 \cite{hns05}. A list of recent studies for
modern societies carried out by both economists and econophysicists, and
which by no means should be considered as exhaustive, includes
Australia \cite{bym06,mah04},
Brazil \cite{cfl98},
China \cite{crt07},
France \cite{qd06},
Germany \cite{qd06},
India \cite{s06},
Italy \cite{cg05,qd06},
Japan \cite{anostt00,fsaka03,i05,s02,s01,sn05},
Poland \cite{dj02,lo04},
United Kingdom \cite{dy01b,h81,qd06,wm04} and
USA \cite{bmm96,s05,cy05,dy01,dy01b,lo04,mc90,wm04,y08}.
The results coming out of these studies are varied. Although most
of them confirm the validity of the Pareto law at higher personal
income data, characterization of the lower individual income
distribution remains disputed. Gaussian, log-normal, gamma,
generalized beta of the second kind, Fisk and Beaman distribution
functions have been used to fit the data, as well as Dagum,
Singh-Maddala and Weibull models
\cite{bj05,bmm96,crt07,h81,k80,l68,lo04,mr07,mc90,qd06}.
Recently the exponential was found to produce a good description for
about 98\% of the population at the lower personal income portion
\cite{bym06,s05,cy05,dy00,dy01,dy01b,dy02,lo04,wm04,y08}.

Disparate interpretations for these distributions have also been
advanced. Many interpretations are basically of statistical nature,
invoking stochastic processes \cite{bym06,dy01,k80,n05,r03,sr01}.
Others attempt to draw analogies from physics. This is the case of
Dr\u{a}gulescu and Yakovenko \cite{dy01,dy01b,y03,y08}, who advanced
an exponential type distribution of personal income analogous to
the Boltzmann-Gibbs distribution of energy in statistical physics,
and Chatterjee et al.\ \cite{ccm04}, who proposed an ideal-gas model
of a closed economic system where total money and number of agents
are fixed.

The purpose of this paper is to study the personal income
distribution of Brazil for approximately the last 30 years.
Here we provide empirical evidence which confirms that Brazil also
follows the Pareto law for the tiny group which constitutes the
high personal income population. The other motivation of this paper
was to try to determine whether or not the exponential is as good a
descriptor for the Brazilian data as it is for the USA. Our results
show that the exponential and, by extension, any function based on
it, turned out to be a very poor descriptor of the lower income
distribution in Brazil. Such a result led us to search for another
simple function capable of describing the individual income
distribution for the majority of the Brazilian population. We propose
here the \textit{Gompertz curve} \cite{g25,kot01,w32} as a good
descriptor for the distribution of the lower income population.
Although the Gompertz curve can be written with two parameters only,
we shall show below that one of them can be linked to a boundary
condition determined by the problem. This effectively leaves only
one parameter to be fitted by the data. Therefore, here we provide
empirical evidence that the personal income distribution in Brazil
reasonably follows the Gompertz curve for the overwhelming majority
of the population.

Our results show that the individual income distribution data in
Brazil from 1978 to 2005 are well described by both the Pareto
law and the Gompertz curve. This time span constitutes virtually
all data for the Brazilian individual income distribution available
in digital form at the time of writing. We have calculated the
parameters of both curves with their uncertainties for all years
in this period, with exception of those when there was no data
collection: 1980, 1991, 1994, 2000 (see Section \ref{data} below). We
also present the Lorenz curves, the Gini coefficients and the
evolution of the Pareto index, that is, the exponent of the Pareto
law, in this time span as well as a comparison of the income share
for the two groups, showing an approximate cycle with roughly a
4 year period. As it happens for other countries, we found
evidence that the lower income population, represented here by a
Gompertz curve, constitutes about 99\% of the Brazilian population,
with the remaining 1\% richest being represented by a Pareto power
law distribution. Similarly to other countries, such results
characterize Brazil as being a well defined two income class
system.

The plan of the paper is as follows. Section \ref{data} presents the
income data of Brazil and discusses how the data reduction necessary
for our analysis was carried out. Some results obtained directly from
the data, such as the Lorenz curves and Gini coefficients are also
shown. Section \ref{model} presents our analytical modeling by
means of the Gompertz curve and Pareto power law complementary
cumulative distribution functions. The results are presented in
Section \ref{results}, where one can find various tables presenting
the fitted parameters and plots showing the linearization of both
the Gompertz and Pareto income regions with their fitted lines, as
well as the evolution of the Paretian component income share relative
to the overall income. Section \ref{conclusion} summarizes and
discusses the results.

\section{The Data}\lb{data}

Personal income data for the Brazilian population is available
in yearly samples called PNAD. This is a Brazilian Portuguese
acronym meaning ``National Survey by Household Sampling.'' IBGE,
the Brazilian government institution responsible for data
collection, formatting and availability, carries out the survey
every September and the data is released usually about one year
later. PNAD data has been systematically available digitally
since 1978, although in 1980, 1991, 1994 and 2000 there was no
data collection and, therefore, there are no PNADs for these
years.  IBGE also has digital PNAD data for 1972, but the file
seems incomplete and without clear labels for each entry. In
addition the 1972 data collection was apparently carried out
by a very different methodology than the one adopted by IBGE
from 1978 onward. For these reasons we considered the 1972
PNAD data unreliable and discarded it from our analysis.

PNAD comprises surveys of about 10\% of households in Brazil.
The released data is made of files with entries for each
surveyed household, providing the total household's income, the
number of people living in, a weight index representing its
proportion to the complete set of households in Brazil, occupation
of those individuals and many other entries which are not relevant
for the present analysis. PNAD is a sampling, not a census, and
the surveyed households' locations in Brazilian territory are
carefully selected by IBGE such that once the weight index is used
the final set should be very close to the complete real set.

The most appropriate procedure to find the personal income from
our data set would be to adopt some sort of ``equivalence scale'',
that is, a tool allowing us to reach conclusions about how the
total income in a household is shared among all of its members.
One way of doing this is to allocate points to each individual
in a household, such that the first adult would have a higher
weight than other persons whose ages are, say, 14 years or older.
Children under the age of 14 would be allocated an even smaller
weight. For instance, the first adult would have a weight of 1
point, additional persons above 14 years would have 0.5 points
and children would be allocated with 0.3 points.  The idea behind
this procedure is to differentiate the household members who
consume, but do not produce income (children, for instance), from
those who do both, but at different levels, and also take into
account the fact that there are goods in a household which are
consumed by several individuals at the same time, like, for
instance, washing-machines, kitchens, etc, and, therefore, a
second adult would not consume as much as the first and would
contribute more in raising the household's well-being. Using
this procedure the income of children under 14 years would be
near zero, even though they share the household's total income.
Equivalized household income would then be obtained by dividing
the total household income by the sum of the points attributed
to the household members \cite{deaton}. 

The major obstacle we faced in implementing such a differentiated
equivalence scale with our data is the fact that the PNADs do not
provide us with enough information to do so. What we have is a list
of the total income in a household and the number of people living
in.
Under these circumstances we adopted an equivalence scale such that
each individual is allocated a weight of 1 point. So, for each PNAD
entry we divided the total income by the number of people living in,
meaning that the household income is equally divided for every
occupant.

As mentioned above, each PNAD household entry has a supplied weight
index corresponding to its relative importance, or representation,
as regards the entire country. This means that although the survey
comprises only a portion of Brazilian households, once we obtain the
income of each individual in a particular home we multiply the
resulting values by this weight in order to obtain the number of
individuals with that particular income in the whole country. Thus,
we end up with tables relating on one side a certain number of
individuals and on the other their respective incomes.

Brazil experienced runaway inflation and hyperinflation for
most of the 1980s and early 1990s, resulting in a series of
currency adjustments where many zeros were ``dropped'' from
time to time and new currency names were adopted each time those
adjustments became effective. Hyperinflation came to an abrupt end
in 1994 when a new and stable currency, called \textit{real} (R\$),
was adopted. This fact required the adoption of a methodology such
that the final data were somehow homogenized, otherwise comparison
of data sets of different years would be problematic. Thus, our
adopted procedure was of normalizing the income values by the average
income of September of each year. In other words, let $x_i'$ be
the \textit{i}th income received on the month of September of a
certain year given in one of the Brazilian currency units legally
adopted in the country when the survey was carried out. Then
$\langle x' \rangle$ is the average income value during the month
of September of that particular year. We may now define the
\textit{normalized individual income} $x_i$ to be the ratio
$x_i = {x_i'}/{\langle x' \rangle}$ so that $x_i$ becomes currency
independent. In this way we were able to produce tables listing the
number of people in terms of currency free income values. This 
allowed us to generate distribution functions relative to the
average personal income in a certain year. This individual average
income does change from year to year, as can be seen in table
\ref{tab1}, where the currency names, exchange rates and the
average individual incomes on September of each year are presented. 
\begingroup
\begin{table*}[!htbp]
\caption{Currencies in Brazil from 1978 to 2005 and the average
         individual income $\langle x' \rangle$ calculated in 
         September of a given year. $\langle x' \rangle$ is converted
         by the exchange rate of September 15th of each year and
         presented in US dollars of that particular day (source:
	 Brazil Central Bank). The hyperinflation period is clearly
	 visible in the evolution of the exchange rate.
	 \label{tab1}}
\begin{center}
\begin{tabular}{crrr}
\hline\noalign{\smallskip}
\textbf{year} & \textbf{currency name - symbol} & \textbf{exchange
rate: 1 US\$ =}& $\mathbf{\langle x' \rangle}$ \textbf{in US\$} \\ 
\noalign{\smallskip}\hline\noalign{\smallskip}
1978 & cruzeiro - Cr\$ &19.05 Cr\$&132.503\\
1979 & cruzeiro - Cr\$ &28.793 Cr\$&122.531\\
1981 & cruzeiro - Cr\$ &105.284 Cr\$&88.201\\
1982 & cruzeiro - Cr\$ &202.089 Cr\$&94.505\\
1983 & cruzeiro - Cr\$ &701.388 Cr\$&56.786\\
1984 & cruzeiro - Cr\$ &2203.96 Cr\$&51.764\\
1985 & cruzeiro - Cr\$ &7461.575 Cr\$&58.039\\
1986 & cruzado - Cz\$  &13.84 Cz\$&91.306\\
1987 & cruzado - Cz\$  &49.866 Cz\$&75.687\\
1988 & cruzado - Cz\$  &326.233 Cz\$&87.276\\
1989 & cruzado novo - NCz\$&3.267 NCz\$&137.183\\
1990 & cruzeiro - Cr\$ &75.54 Cr\$&161.516\\
1992 & cruzeiro - Cr\$ &5775 Cr\$&111.906\\
1993 & cruzeiro real - CR\$ &111.1 CR\$&126.547\\
1995 & real - R\$      &0.953 R\$&213.188\\
1996 & real - R\$      &1.019 R\$&228.299\\
1997 & real - R\$      &1.094 R\$&221.687\\
1998 & real - R\$      &1.181 R\$&213.902\\
1999 & real - R\$      &1.898 R\$&133.631\\
2001 & real - R\$      &2.672 R\$&110.857\\
2002 & real - R\$      &3.342 R\$&97.532\\
2003 & real - R\$      &2.923 R\$&122.815\\
2004 & real - R\$      &2.891 R\$&148.468\\
2005 & real - R\$      &2.294 R\$&221.787\\
\noalign{\smallskip}\hline
\end{tabular}
\end{center}
\end{table*}
\endgroup

Our next step was then to divide the data in bins inasmuch as
most data is clumped towards low income values. The data
binning methodology adopted here is the standard one used for
problems involving power law determination \cite{n05} and which
was previously used by these authors to derive the Zipf law for
Brazilian cities \cite{nm}. The method consists of taking
logarithmic binning such that bins span at increasing larger
intervals and every step is 10\% larger than the previous
one. This is accomplished according to the rule below,
\be x_j= 1.1^{(j-1)}x_{\mathrm{min}}.\lb{rule} \ee
By following this procedure we were able to create for each year
a sample of $n$ observed values such that,
$$
  \{x_j\}: (j=1,\ldots,n), \, (x_1=x_{\mathrm{min}}), \,
  (x_{\mathrm{min}}\approx 0.01), \, (n \approx 100).
$$
The purpose of this methodology is to achieve a sharp
decrease in the statistical fluctuations in the tail due to the
fact that bins with far smaller number of observed values, prevalent
in the tail of the distributions, are prone to large fluctuations.
This effect has the potential of creating a serious bias in the
determination of the parameters by least square fitting \cite{n05}.
To counteract this problem, it is known that an appropriate
logarithmic binning is very effective at severely reducing the
uneven variation in the tail, which means that the possible bias
in the parameter determination by least square fitting \cite{g04}
is, therefore, strongly reduced.

After the steps described above were taken we were able to
obtain cumulative probabilities by calculating the number of
individuals whose income goes up to certain values and
dividing this value by the total number of individuals. The final
results are shown in figures \ref{fig1} and \ref{fig2}, where
complementary cumulative probabilities are plotted against normalized
income for each year of the studied time span. It is clear
from these graphs that there are enough points to form an almost
continuous and smooth curve. Therefore, from now on we will change
the discrete variable $x_j$ to the continuous independent variable
$x$ representing the normalized individual income values.
\begin{figure*}
\epsfysize=16cm
\begin{center}
\rotatebox{-90}{\epsffile{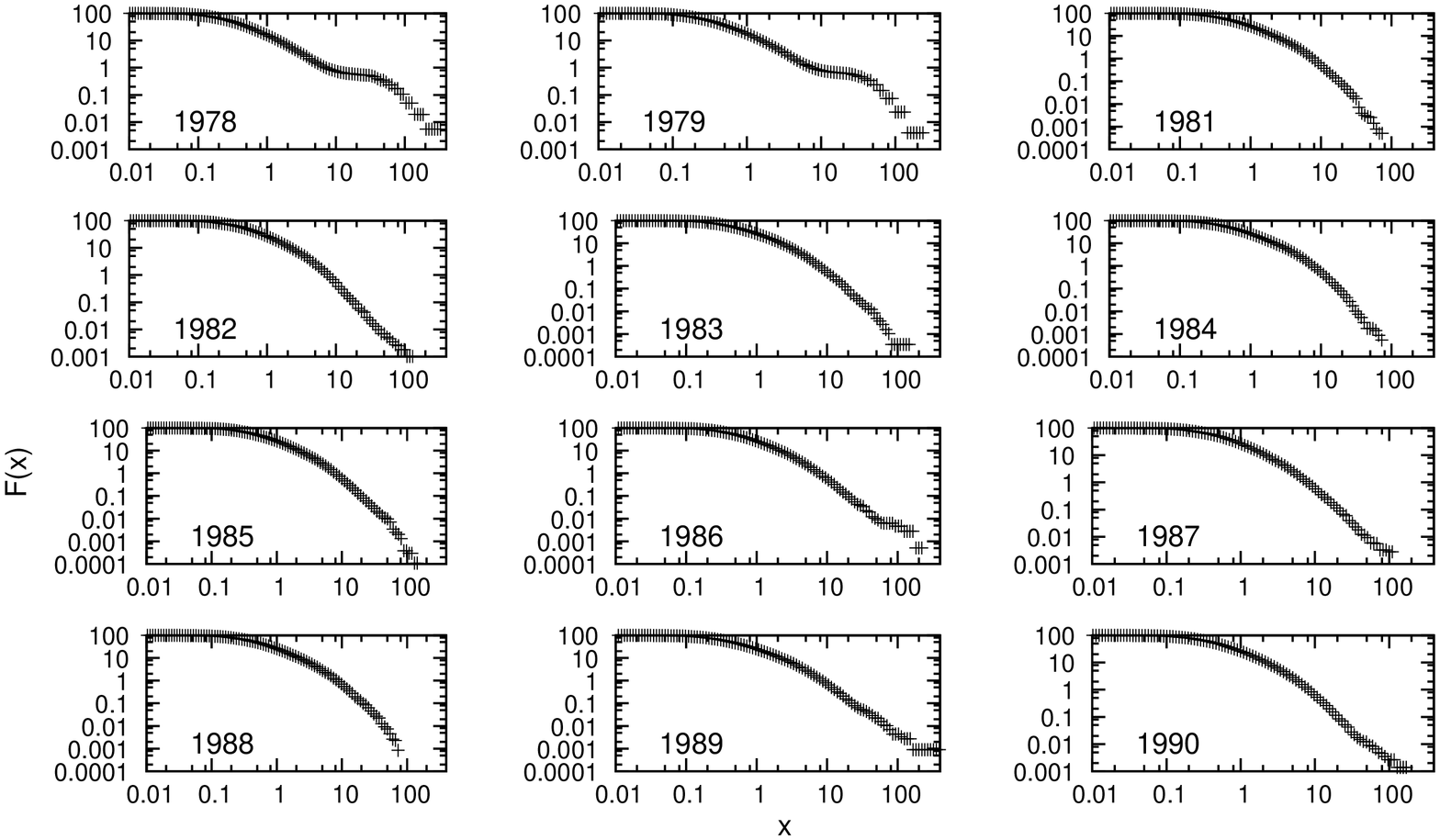}}
\end{center}
\caption{Graph of the complementary cumulative probability of
         individual income $F(x)$ plotted
	 against the normalized individual income $x$ 
	 for the month of September of each year in the time span
	 of this study. Although Brazil experienced runaway inflation
	 and hyperinflation from 1981 to 1993 the plots show a
	 remarkable similarity during and after this period. The
	 major differences stem from the plots of 1978 and 1979, just
	 prior to Brazil's great inflationary period. Due to the
	 absence of reliable digitalized data before 1978 we were
	 unable to ascertain whether or not these years were the last
	 ones of a qualitatively different era regarding the income
	 distribution in Brazil, which was then possibly terminated
	 by the inflationary period.\label{fig1}}
\end{figure*}
\begin{figure*}
\epsfysize=16cm
\begin{center} \rotatebox{-90}{\epsffile{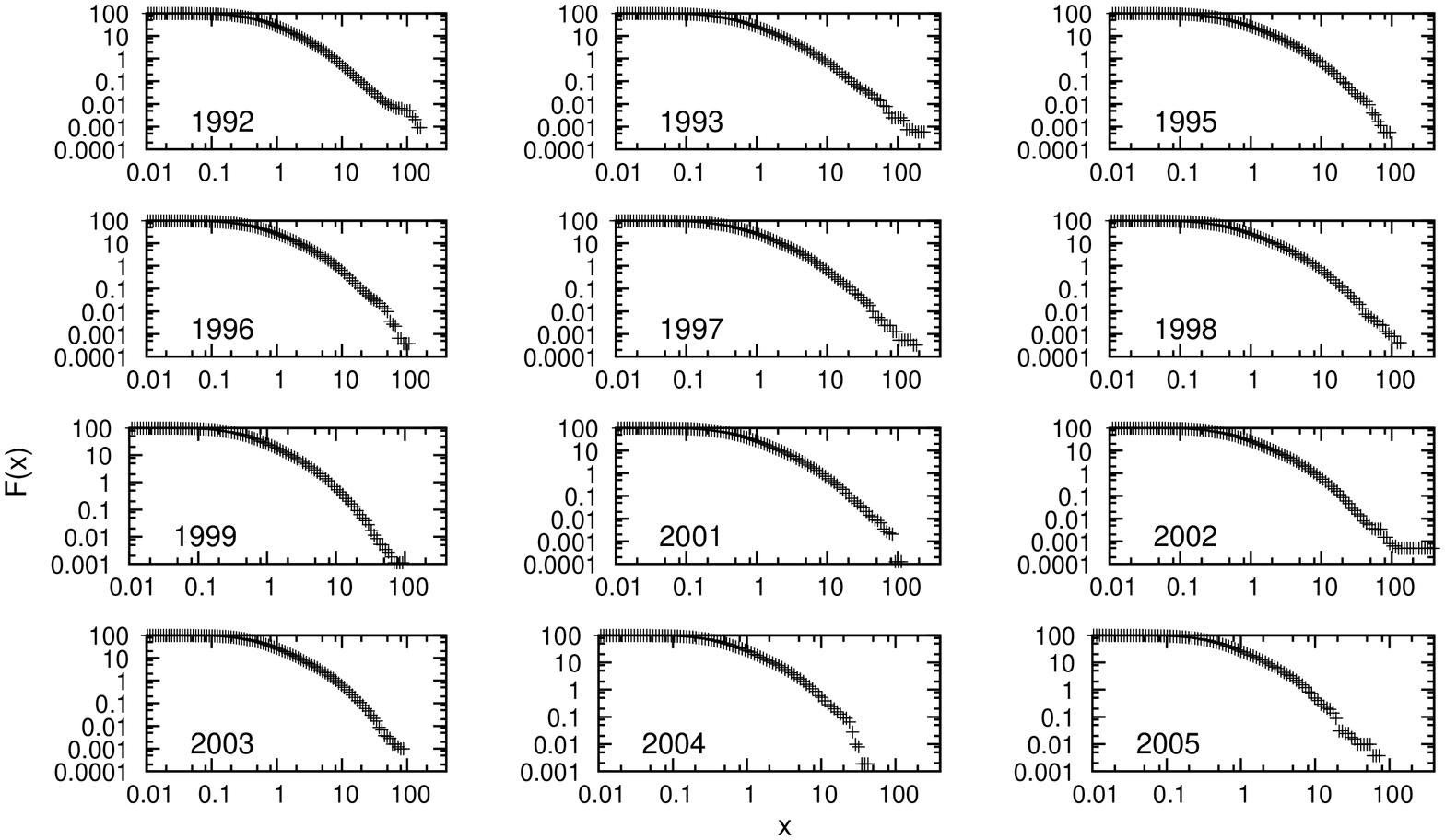}} \end{center}
\caption{Continuation of figure \ref{fig1} showing the complementary
         cumulative individual income distribution $F(x)$ against
	 the normalized individual income $x$ in Brazil from 1992 to
         2005.\label{fig2}}
\end{figure*}

The data obtained with the procedures outlined above allowed us to
calculate the so-called \textit{Lorenz curve} \cite{k80,l05},
which measures the degree of inequality in income distribution,
by setting the maximum income value to 100\% and then calculating
the percentage of individuals who receive
certain percentage of the maximum income. Figures \ref{fig3} and
\ref{fig4} show the Lorenz curves for Brazil from 1978 to 2005.
\begin{figure*}
\epsfysize=16cm
\begin{center} \rotatebox{-90}{\epsffile{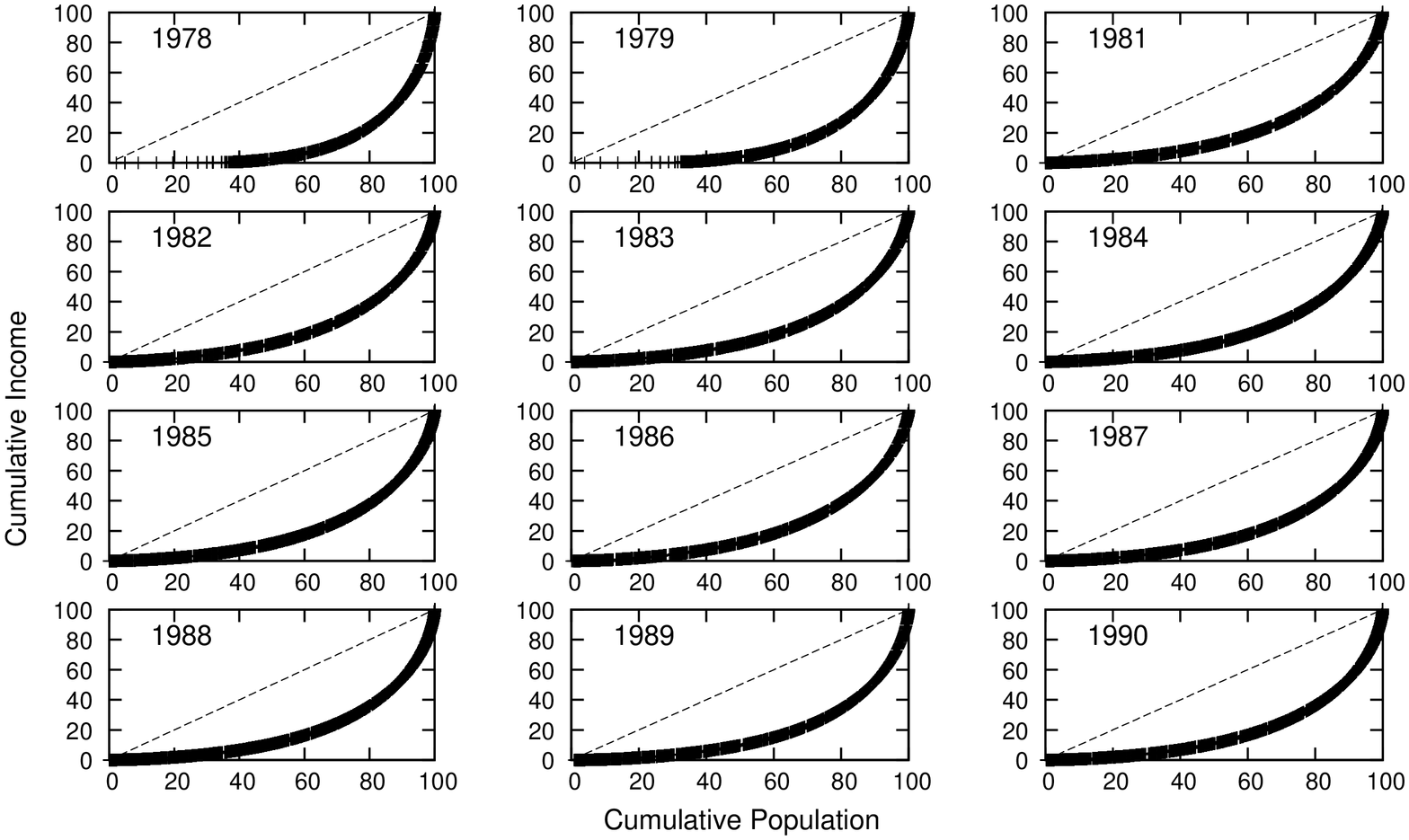}} \end{center}
\caption{The Lorenz curves for the individual income distribution of
         Brazil in the month of September of the respective year. The
         x-axis plots the \% of individuals whereas the y-axis is the
         \% of total income.\label{fig3}}
\end{figure*}
\begin{figure*}
\epsfysize=16cm
\begin{center} \rotatebox{-90}{\epsffile{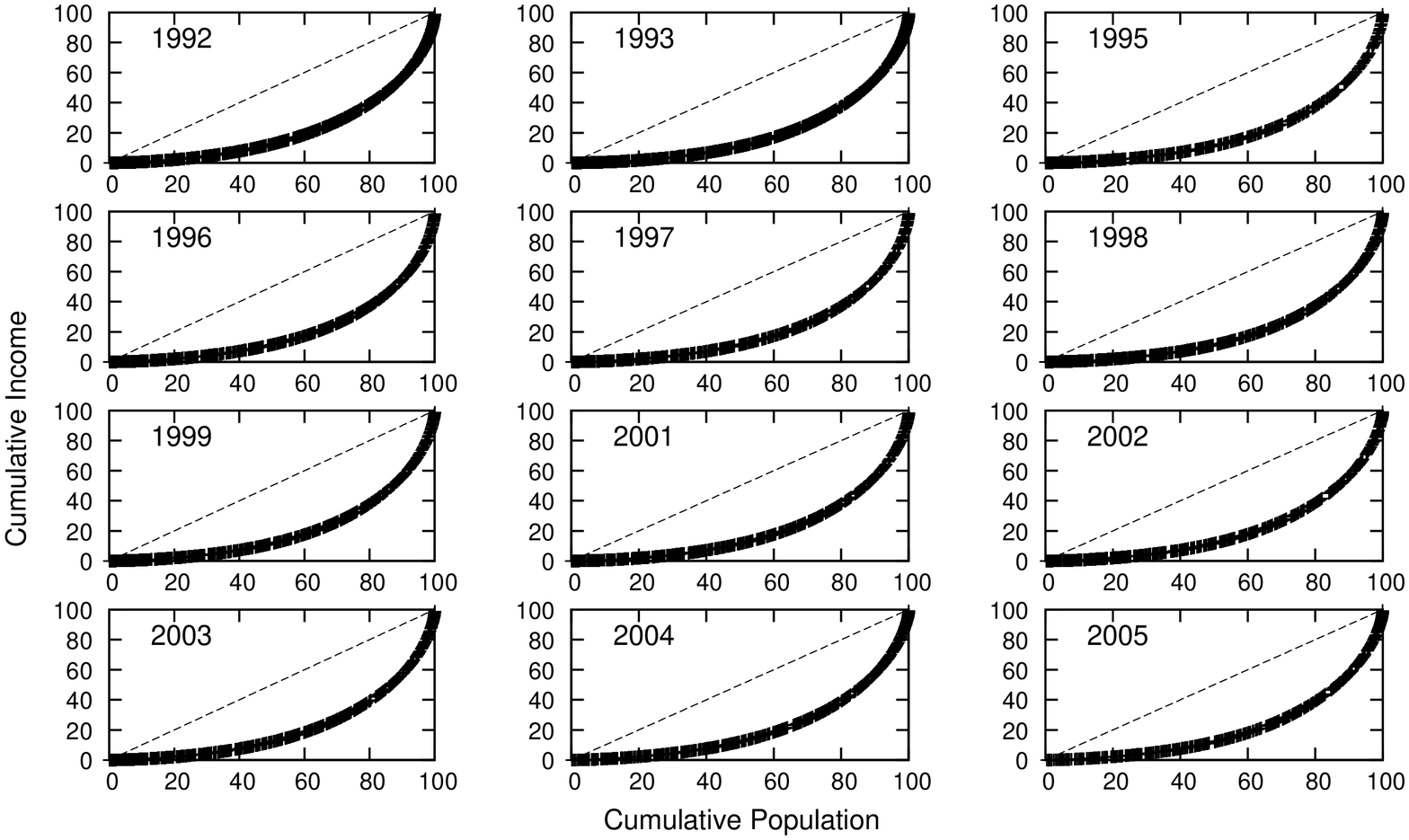}} \end{center}
\caption{Continuation of figure \ref{fig3} showing the Lorenz curves
         of the income distribution in Brazil from 1992 to 
         2005.\label{fig4}}
\end{figure*}

\begin{figure*}
\epsfysize=13cm
\begin{center} \rotatebox{-90}{\epsffile{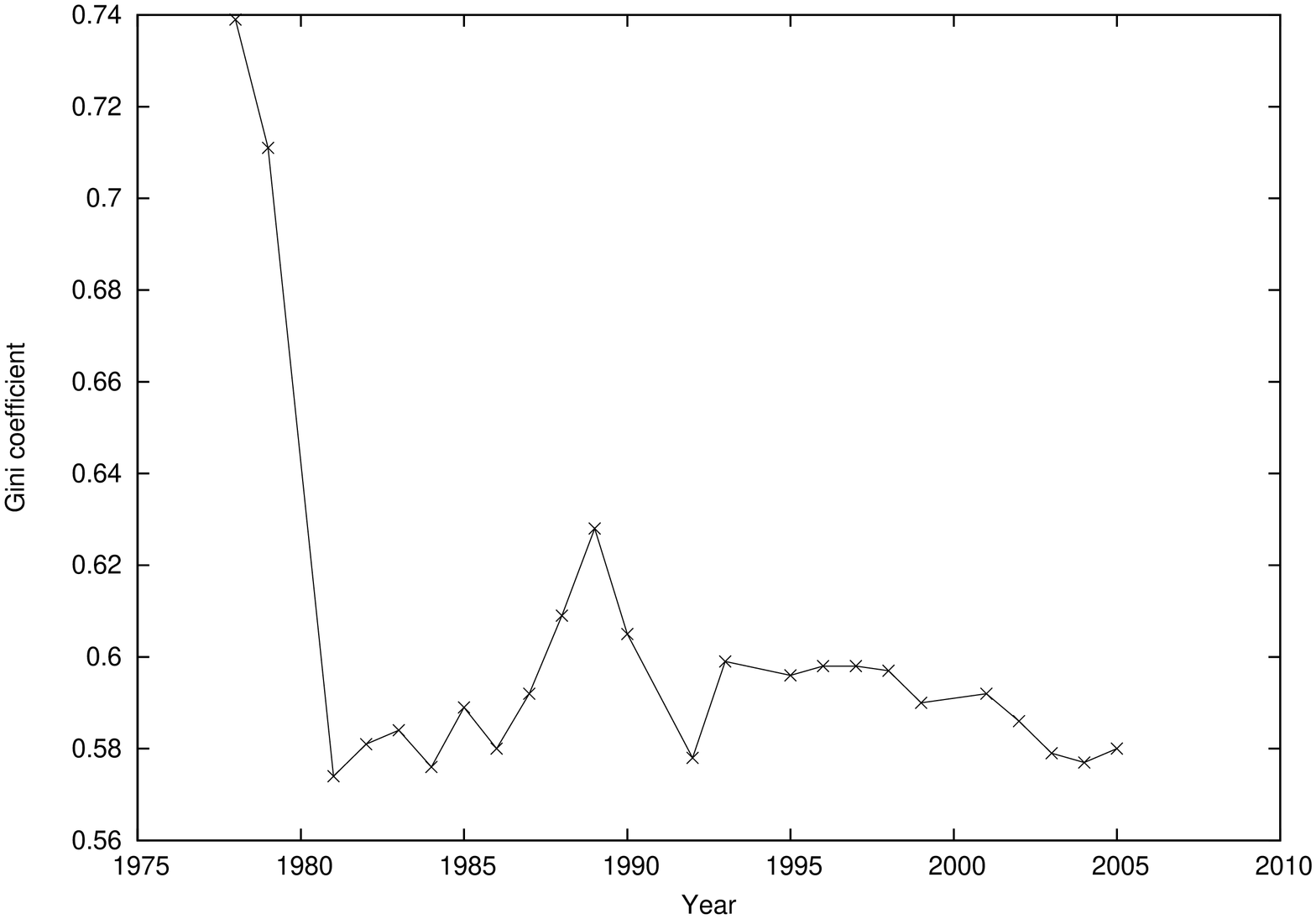}} \end{center}
\caption{This figure shows the evolution of Brazilian Gini coefficient
         for most of the last three decades. The values shown in this
	 plot are presented in table \ref{tab4a}.\label{fig5}}
\end{figure*}

Once the points forming the Lorenz curves had been calculated we were
able to obtain the correspondent \textit{Gini coefficients}
\cite{g12,g13,g21,k80}, which measure the inequality of the income distribution.
This was done by numerically calculating the area below the Lorenz
curves. Figure \ref{fig5} shows the results.

\section{Modeling the Individual Income Distribution}\lb{model}

Anyone attempting to familiarize oneself with the recent literature
in econophysics will see that when physicists try to solve problems
traditionally dealt with by economists they do so via a different
perspective. That, of course, will be no different for the income
distribution problem. We therefore believe to be fruitful to expose
our viewpoints about how to approach the income distribution problem
at the very beginning of our discussion. So, this section will start
by outlining our modeling perspective and how it differs from the
traditional approach followed by economists.

The first aspect worth mentioning is that economists often fit
their income distribution data by means of complex single functions
with as many
parameters as necessary \cite{bj05,bmm96,crt07,dj02,mc90,qd06}.
Fitting the whole dataset with single functions with four or more
parameters may produce a better data fit, but the drawback is that
this kind of fitting does not give a better insight into the problem.
The paramount objective of any physical modeling is to find the
differential equations which describe the observed empirical pattern
and, therefore, data fitting is only the very first step into that
direction and must be made bearing in mind Occam's razor, which in
this case means using simple functions with as little parameters as
possible. Theoretical assumptions of economic nature must be built
into the differential equations and not in the empirical curves.
Therefore, the relationships among the parameters should be a
result of the dynamics of the model determined by the solutions of
the differential equations. Using as many parameters as necessary
in complicated functions which do not originate from some sort of
dynamical analysis is not a promising approach to the income
distribution problem because it will make the task of finding the
underlying differential equations even more difficult, if not
impossible. Perhaps this is one of the reasons
why the approach of conventional mainstream economics to the
personal income distribution problem has made little progress
since Pareto's time towards developing a dynamical theory
connecting personal income generation and economic growth in the
sense of Sraffa \cite{s60}, as pointed out by Gallegati et al.\
\cite{gklo06}. Thus, simple functions with as few parameters as
possible which, at the same time, offer a reasonable agreement
with the data are certainly much more preferable.

The second point is that there is a tendency among a sizable 
number of economists of following an axiomatic and mathematically
guided approach to their problems
as opposed to the empirically
guided paths usually taken by physicists. The major trouble of
approaching a problem guided almost exclusively by logic is that
this often leads to paradoxical situations, where it is possible
to deductively arrive at apparently sound conclusions, which at
the same time are entirely unsound empirically -- here
Aristotelian physics comes to mind as an example. The empirically
sound path means starting and staying as close to the real data
as possible when studying any problem of economic nature and
avoiding as much as possible any kind of a priori assumption. This
is especially true during the infancy of a new area of study.
Examples of successful theories which did not follow this path
are exceedingly rare, even within physics. That does not mean we
dismiss the power of theoretical reasoning, but even 20th century
theoretical physics is strongly anchored upon very solid empirical
foundations. For this reason we believe that research in
econophysics must always carefully consider the real data in
order to avoid at all costs hypothetical, often anti-empirical,
a priori assumptions. For econophysics to succeed it must not
repeat the fatal traps of conventional neoclassical economics,
which is based on too many anti-empirical assumptions, resulting
in all too often compromised results
\cite{jpb08,g06,gklo06,k01,keen03,ks06,mh04,mc00,mc04,mc05,mc07,o97}.

As a third point, it was mentioned above that Dr\u{a}gulescu and
Yakovenko \cite{dy01,dy01b,y03} proposed an exponential type
distribution of personal income analogous to the Boltzmann-Gibbs
distribution of energy in statistical physics under the motivation
that ``in a closed economic system money is conserved'' \cite{dy00}.
Similarly Chatterjee et al.\ \cite{ccm04} advanced an ideal-gas
model of a closed economic system where total money and number of
agents are fixed such that ``no production or migration occurs and
the only economic activity is confined to trading'' \cite{ccm04}.
Those results led to criticisms made by Gallegati et al.\
\cite{gklo06} who argued that industrialized economies are not a
conservative system, meaning that ``income is not, like energy in
physics, conserved by economic processes''. This occurs because
although transactions, that is, exchanges are conservative,
``capitalist economies are (...) characterized by economic growth.
And growth occurs because production produces a net physical surplus''.
Ref.\ \cite{gklo06} concludes by stating that ``models which focus
purely on exchange and not on production cannot by definition offer
a realistic description of the generation of income in the
capitalist, industrialized economies''.

Gallegati et al.\ \cite{gklo06} may have a point regarding
the development of a dynamical theory of production. However, 
the focus of the approach made by physicists on the personal
income distribution characterization problem has not been on this
dynamical theory, which is obviously necessary, but has not
yet been developed. So far, econophysicists have been mainly
focused on the more modest aim of finding
good analytical descriptors of the individual income distribution,
not only for the very rich where the Pareto law is valid, but for
the whole society. On this point the proposal of an exponential
distribution is without any doubt a step forward since it seems
to produce good agreements with the data of some countries and is
a simple function, with one parameter only. Therefore, if the
exponential function does not produce a good fit for the income
data of Brazil (see below) we are entitled to ask whether or not
it is possible to find another function with one, or two parameters
at most, which could produce a good data fit for the Brazilian
data and, perhaps, could also be useful for fitting the income
data of other countries.

As a final conceptual point, we should mention that in recent
econophysics literature the words ``income'' and ``wealth''
have been used indistinctively. We believe this to be inappropriate.
In this article income is used as a generic term for anything gained
by an individual in a specific period of time, usually monthly or
annually. It can be wage, pension, government grant, the revenue
obtained from property or investment like rent or dividends, etc.
However, we believe that income should not be confused with wealth,
because although these two concepts are related, wealth is the result
of saved, or accumulated, income, often inherited. In other words,
income is a flux, an inflow of value that an individual receives, or
earns, at a specific time interval which, if accumulated, may become
wealth. In turn, the investment of wealth in property, shares, etc,
generates income as rent, dividends, etc. The empirical findings that
led to Pareto law were mostly derived from personal income data,
although it appears reasonable to suspect that the personal wealth
distribution should also follow a power law for those individuals
with high wealth. 

\subsection{Basic Equations}

Let $\mathcal{F}(x)$ be the {\it cumulative distribution
function of individual income}, or simply {\it cumulative income
distribution}, which gives the probability that an individual
receives an income less than or equal to $x$. It follows from
this definition that the \textit{complementary cumulative income
distribution} $F(x)$ will then give the probability that an
individual receives an income equal to or greater than $x$.
Clearly $\mathcal{F}(x)$ and $F(x)$ are related by the following
expression,
\be \mathcal{F}(x)+F(x)=100,
    \lb{ff}
\ee
where we have assumed the maximum probability as being equal to
100\%. If both $\mathcal{F}(x)$ and $F(x)$ are continuous and
have continuous derivatives for all values of $x$, this means that,
\be d\mathcal{F}(x)/dx = f(x), \; \; \; \; dF(x)/dx=-f(x),
    \lb{c}
\ee
and
\be \int_0^\infty f(x)\:dx=100. \lb{norm}
\ee
Here $f(x)$ is the {\it probability distribution function of individual
income}, defined such that $f(x)\,dx$ is the fraction of individuals
with income between $x$ and $x+dx$. This function is also known
as {\it probability density}, but from now on we will call it
simply as {\it probability income distribution}. The equations above
lead to the following results,
\be \mathcal{F}(x) - \mathcal{F}(0) = \int_0^x f(w) \: dw,
    \lb{3}
\ee
\be F(x) - F(\infty) = \int_x^\infty f(w) \: dw.
    \lb{4}
\ee

Although we found in our data a non-negligible number of individuals
who earned nothing when the sampling was carried out, zero income
values do not have a weight in the income distribution function
and, therefore, it seems reasonable to assume those results to be
of a transitional nature and dismiss them from our analysis by
assigning zero probabilities. Similarly, very rich people are made
of very few individuals such that their probabilities tend to zero.
Note, however, that these two situations are limiting cases and
should only be considered as true within the uncertainties of our
measurements. Therefore, it follows from this reasoning that the
boundary conditions below should approximately apply to our problem,
\be \left\{ \begin{array}{lclcl}
    \mathcal{F}(0) & = & {F}(\infty) & \cong & 0, \\
    \mathcal{F}(\infty) & = & {F}(0) & \cong & 100. 
            \end{array}
    \right.
    \lb{condi1}
\ee

\subsection{Two Parts for the Income Distribution}

As discussed above, our approach implies searching for simple
functions to describe the income distribution. Therefore we
shall divide this distribution in two distinct parts, one for
the very rich and the other for the overwhelming majority of
the population. To establish the notation, when divided that
way the complementary cumulative distribution function of
the individual income will be written as follows,
\be F(x)= \left\{ \begin{array}{ll}
          G(x), & \; \; ( \: 0 \le x < x_{\sty t}), \\
	  P(x), & \; \; (x_{\sty t} \le x \le \infty), \\
	  \end{array}
	  \right.
          \lb{disto}
\ee
where $x_{\sty t}$ is the \textit{transitional income value}
marking the transition between the two components of the
income distribution. Then the cumulative distribution will
be given by,
\be \mathcal{F}(x)= \left\{ \begin{array}{ll}
          \mathcal{G}(x), & \; \; ( \: 0 \le x < x_{\sty t}), \\
	  \mathcal{P}(x), & \; \; (x_{\sty t} \le x \le \infty), \\
	  \end{array}
	  \right.
          \lb{disto1}
\ee
and the probability density yields,
\be f(x)= \left\{ \begin{array}{ll}
          g(x), & \; \; ( \: 0 \le x < x_{\sty t}),  \\
          p(x), & \; \; (x_{\sty t} \le x \le \infty).  \\
	  \end{array}
	  \right.
          \lb{distro2}
\ee

\subsection{The Pareto Law}

It is a well known empirical fact that the richest portion of many,
perhaps most, populations follows a {\it Pareto power law} of the
form,
\be P(x) = \beta \; x^{\displaystyle - \alpha},
    \label{pareto}
\ee
where $\alpha$ and $\beta$ are positive constants. The parameter
$\alpha$ is known as \textit{Pareto index} or just the \textit{
fractal dimension} of the distribution, if we adopt the modern
language of fractals \cite{mandelbrot,mh04}. This law is valid
only for the region of \textit{high personal income}, starting
at $x=x_{\sty t}$ and going up to the maximum value obtained in
the observed dataset. As we shall show below our data presents
compelling evidence that the Pareto law is valid in Brazil.

It is well known that if the complementary cumulative distribution
is a power law, the probability distribution is also a power law.
Therefore, the Paretian part of the income distribution of the
Brazilian population has a probability density given by the
following expression,
\be p(x)= \alpha \; \beta \; x^{^{\scriptstyle -(1+\alpha )}}.
    \lb{pareto1} 
\ee
It clearly follows from this equation that $p(\infty)=0$. 

\subsection{The Lower Income Region}

\subsubsection{The Exponential}

The first obvious thing to do with our data in the lower income
region was to follow the proposal of Ref.\ \cite{dy01} and try
an exponential fit. Surprisingly, however, the results were
not good. The semi-log plot clearly did not linearize our data,
something that could only be achieved by removing the values due
to very low income. Figures \ref{exp1} and \ref{exp2} show plots
where we have attempted to fit the exponential to the Brazilian
data and a simple visual inspection shows the inadequacy of this
function to describe the observed data points. Since other
functions like the Gaussian or the Boltzmann-Gibbs are also
derived from the exponential, these graphs were enough to
convince us to dismiss all functions based on a simple exponential
as viable fits for the Brazilian data. We then started searching
for other ways of representing the Brazilian income distribution,
especially at very low income values.
\begin{figure*}
\epsfysize=16cm
\begin{center} \rotatebox{-90}{\epsffile{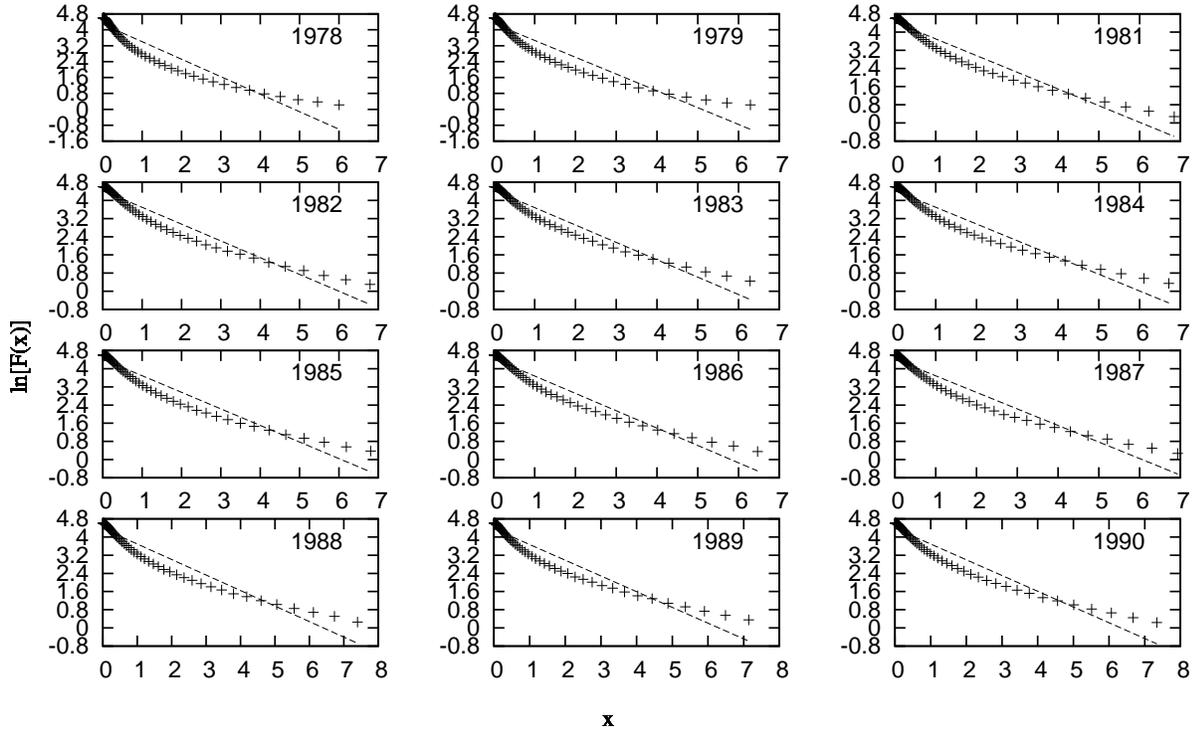}} \end{center}
\caption{These graphs show the exponential fit for the lower region
         of the income distribution. Clearly the exponential is not
	 a good representation for the Brazilian income data.\label{exp1}}
\end{figure*}
\begin{figure*}
\epsfysize=16cm
\begin{center} \rotatebox{-90}{\epsffile{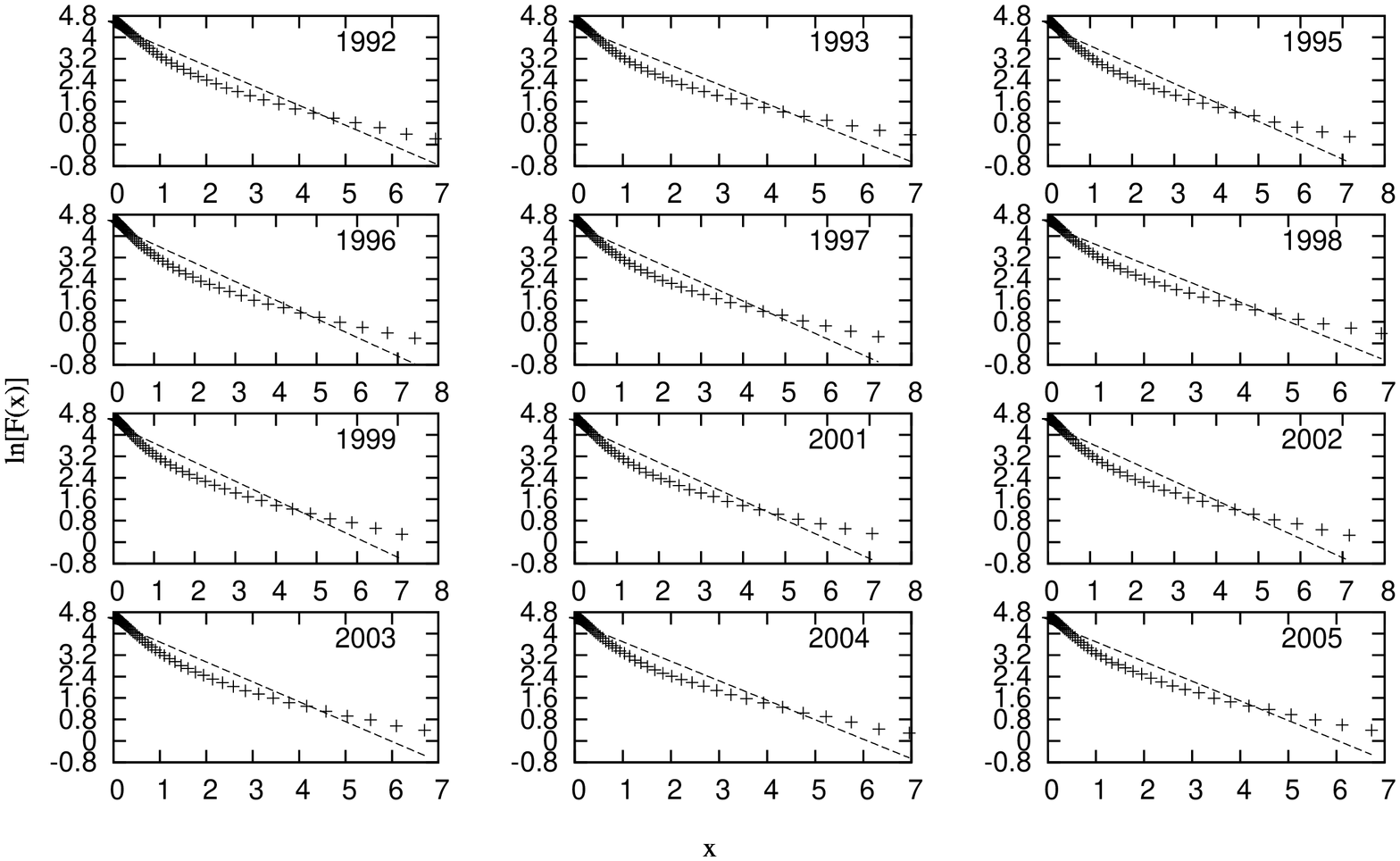}} \end{center}
\caption{Continuation of figure \ref{exp1} showing how poor
         representation is the exponential for the complementary
	 cumulative income distribution data in Brazil.\label{exp2}}
\end{figure*}

\subsubsection{The Gompertz Curve}

In the process of searching for a simple function capable of
representing our dataset we realized that the plot itself suggested
taking the second logarithm of the complementary cumulative
distribution. When doing so the data tended to follow a straight
line, a result which immediately suggested adopting the {\it
Gompertz curve} \cite{w32} to model the complementary cumulative
income distribution of Brazil. This curve may be written as follows, 
\be G(x)= e^{ \displaystyle e^{(A-Bx)} },
    \label{gomp}
\ee
where $A$ and $B$ are positive constants. Section \ref{conclusion}
below presents further discussions about this function.

The definition of cumulative distribution and its complement allow us
to find the Gompertzian probability density income distribution of the
Brazilian population. It can be written as follows,
\be g(x)= B \; e^{(A-Bx)}
          \; e^{ \displaystyle e^{(A-Bx)}}.
    \lb{gomp1}
\ee

Therefore, as mentioned above, in what follows it will become
clear that our data presents compelling evidence that the
complementary cumulative individual income distribution in Brazil
has two distinct components represented by a Gompertz curve and
the Pareto power law, situation which, similarly to other
countries, {\it characterize Brazil as having a well defined
two income class system} as far as individual income is
concerned.

Both equations (\ref{pareto}) and (\ref{gomp}) can be
linearized and, therefore, the unknown parameters can be
obtained by linear data fitting. However, the boundary
conditions (\ref{condi1}) allow us to find the theoretical
value for $A$ and $g(0)$. These results may be written as
follows,
\be e^{ \displaystyle e^{A}}= G(0), \; \; \; \Leftrightarrow
    \; \; \; A=\ln \left\{ \ln \left[ \, F(0) \right] \right\} =
    1.53,
    \lb{A}
\ee
\be g(0)=461 \: B.
    \lb{g0}
\ee
The equations above are just different ways of expressing the
boundary condition due to zero income individuals data. The
fitting should produce values for $A$ which will probably
fluctuate around its theoretical result above. Finding the
extent of these fluctuations is one of our goals, since they
should indicate how much the approximations given by equations
(\ref{condi1}) are valid. Nevertheless, it is an advantageous
feature of our modeling to know beforehand one of the four
parameters. As we shall see below, $\beta$ can be determined by
either data fitting or normalization, a fact which effectively
leaves only two parameters, $\alpha$ and $B$, to be determined
entirely by data fitting.

\subsection{Continuity Across the Gompertz-Pareto Regions}

It is desirable to investigate whether or not the cumulative income
distribution remains continuous across the transition between the
Gompertz and Pareto regions. For this continuity to occur all
parameters should obey the constraint equation $G(x_t)=P(x_t)$,
that is,
\be 
   e^{ \displaystyle e^{(A-Bx_{\sty t})} }=
   \beta \; x_{\sty t}^{\displaystyle - \alpha}.
  \lb{vinc}
\ee
In addition, should the usual normalization of the probability
distributions between the two regions possibly hold, the following
condition will need to be satisfied,
\begin{eqnarray}
 \int_0^\infty f(x)\,dx = \int_0^{x_{\sty t}}  B \; e^{(A-Bx)}
 \; e^{ \displaystyle e^{(A-Bx)} } dx + \nonumber \\
  +  \int_{x_{\sty t}}^\infty  \alpha \; \beta \; x^{^{\scriptstyle
  -(1+\alpha )}} dx=100.
 \lb{norm1}
\end{eqnarray}
It is straightforward to show that the normalization above
together with the boundary conditions (\ref{A}) lead to the
same constraint equation (\ref{vinc}). It is also simple to
verify that the constraint equation above can be solved once
$\alpha$, $\beta$ and $B$ are determined by fitting, albeit
finding $x_t$ from equation (\ref{vinc}) requires the use of
numerical methods. Nevertheless, our preference is to
determine $x_t$ directly from the observed data, leaving the
remaining parameters to be obtained by a mixture of data
fitting and normalization. 

\subsection{Exponential Approximation of the Gompertz Curve}

We can derive a convenient approximation for the Gompertz curve
(\ref{gomp}) when it nears the Pareto region, i.e., for large values
of $x$. In this case the term $Bx$ dominates over the parameter $A$
and equation (\ref{gomp}) reduces to $G(x) \approx e^{\displaystyle
e^{-Bx}}$. If we now define a new variable $z=e^{-Bx}$, then large 
values of $x$ imply small values of $z$ and the following Taylor
expansion holds:
\be e^z = 1+ z+z^2/2+z^3/6+\ldots \; (z<1).
    \lb{ex}
\ee
In view of this we may write the following approximation,
\be G(x)= e^{ \displaystyle e^{(A-Bx)} } \approx 1+  e^{-Bx} \; \; \;
    (\mbox{for $Bx > A$ and $e^{-Bx} < 1$}).
    \lb{gomp-exp}
\ee
This result means that the Gompertz curve reduces to the exponential
function when the personal income $x$ is large enough. It also means
that the Gompertz curve allows us to have one of its parameters as a
boundary condition for the zero income situation at the same time as
having an exponential feature for larger incomes. In addition,
the probability income distribution as given by equation (\ref{gomp1})
can also be similarly approximated, yielding,
\begin{eqnarray} g(x) & = & B \; e^{(A-Bx)}
          \; e^{ \displaystyle e^{(A-Bx)}} \nonumber \\
	  & \approx & B \; e^{-Bx} \; \;
          (\mbox{for $Bx > A$ and $e^{-Bx} < 1$}).
    \lb{gomp1-exp}
\end{eqnarray}

Note that the approximation above means leaving the very low
income data out of our analysis, which in turn reduces our
problem to the exponential fit, as proposed in Ref.\ \cite{dy01}.
A simple visual inspection of figures \ref{exp1} and \ref{exp2}
shows that the data seems to be fairly represented by an
exponential if we remove the very low income dataset $(x \le 2)$.
This feature may explain why the exponential is such a poor
representation of our income data. Brazil is notoriously
a very unequal country in terms of income distribution and,
therefore, our data tend to clump towards low income values.

Finally, the approximations (\ref{gomp-exp}) and
(\ref{gomp1-exp}) also mean that the exponential
and the Gompertz curve are not very dissimilar to one another
in terms of being good representations of the non-Paretian part
of the individual income distribution. So, \textit{the case for
the Gompertz curve is made on the grounds of a better data fit},
especially considering  the very low income values that are
strongly represented in the Brazilian income dataset, and its
possible interpretation as a growth curve in the context of
attempting to connect personal income with industrial production
and economic growth (see Section \ref{conclusion} below).

\subsection{Average Income}

The mean income of the whole population may be written as follows,
\begin{eqnarray}
  \langle x \rangle & = &\frac{\int_0^\infty x \: f(x) \: dx}{\int_0^\infty
          f(x) \: dx} = \frac{1}{100} \left[ 
        \int_0^{x_{\sty t}} x \: B \: e^{(A-Bx)} \; e^{ \displaystyle
	e^{(A-Bx)}} dx \; + \right. \nonumber \\
	& + & \left. \lim_{x_{_{\mbox{\tiny max}}} \rightarrow
	\infty} \int_{x_{\sty t}}^{x_{_{\mbox{\tiny max}}}} x \: \alpha
	\: \beta \: x^{-(1+\alpha)} dx \right]. 
  \lb{avg}
\end{eqnarray}
The solution of the last integral on the right hand side yields,
\begin{eqnarray} & \lim_{x_{_{\mbox{\tiny max}}} \rightarrow \infty} &
    \int_{x_{\sty t}}^{x_{_{\mbox{\tiny max}}}} x \: \alpha \: \beta \:
    x^{-(1+\alpha)} dx = \nonumber \\ 
    = & \lim_{x_{_{\mbox{\tiny max}}} \rightarrow \infty} &
    \left\{ \frac{\alpha \: \beta}{(1-\alpha)}
    \left[ {x_{_{\mbox{\tiny max}}}}^{(1-\alpha)} -
    {x_{\sty t}}^{(1-\alpha)} \right] \right\}.
    \lb{limit}
\end{eqnarray}
Clearly this limit will only converge if the Pareto index is
bigger than one. Possible non finite averages may happen with
power laws as discussed in Ref.\ \cite{n05}. Indeed, datasets
of finite sizes will produce a finite average since we can take
$x_{_{\mbox{\tiny max}}}$ as being the maximum dataset value
and cut off this integral above some upper limit. Nevertheless,
this is not the case of income distribution because although
there are extremely rich individuals, if we make more
measurements and generate a larger dataset we will eventually
reach a value of $x$ such that the chance of getting an
even larger value will indeed become zero, since even super-rich
individuals do not receive an infinite income and their numbers are
finite. In other words, as we go to larger and larger individual
income datasets our estimate of $\langle x \rangle$ will
\textit{not} increase without bound. We therefore can conclude that
the condition $\alpha > 1$ is an empirically necessary requirement
for the Pareto law to hold, which is just another way of stating
that the boundary condition $F(\infty) \cong 0$ is empirically
sounding. In such a case equation (\ref{avg}) reduces to an
expression which may be written as below,
\be  \langle x \rangle = \frac{1}{100} \left[
      \mathcal{I}(x_t)
      + \frac{\alpha \: \beta}{(\alpha -1)}
      {x_{\sty t}}^{(1-\alpha)} \right], \; \; \; (\mbox{for
      $\alpha > 1$}),
      \lb{avg2}
\ee
where $\mathcal{I}(x)$ is given by the following, numerically
solvable, integral,
\be \mathcal{I}(x)
    \equiv \int_0^x w \, g(w) \, dw=
    \int_0^x w \: B \: e^{(A-Bw)} \;
    e^{ \displaystyle e^{(A-Bw)}} dw. 
    \lb{I}
\ee

\section{Results}\lb{results}

\subsection{Parameters of the Gompertz Curve}

To determine $A$ and $B$ we carried out a least squares fit
since in this region the dataset does not exhibit large
fluctuations which can cause large fitting bias, as
discussed in Goldstein et al.\ (2004). However, to do so we
first need to find $x_{\mathrm{gmax}}$, that is, the maximum
value of $x$ that marks the end of the Gompertz region. The
boundary conditions (\ref{condi1}) and (\ref{A}) imply
$A=1.53$ and, therefore, we assumed that the end of the
Gompertz region is reached when a value for $x$ is found
such that the straight line fit of $\left\{ \ln \left[ \ln
G(x) \right] \right\}$ produces $A=1.5\pm0.1$. By following
this methodology we were able to determine the specific value
of $x_{\mathrm{gmax}}$ for our dataset and fit the Gompertz
curve. Plots are shown in figures \ref{gomp1f} and \ref{gomp2}
and the results are summarized in table \ref{t-gomp} where
one can verify that the result $A=1.54\pm0.03$ encompasses
the whole period under study, that is, from 1978 to 2005.
Hence, in the time period of our analysis $A$ varies no more
than 2.6\% from its boundary value given in equation (\ref{A}).
Regarding the other parameter, the results are also stable
from 1981 to 2005. However, $B$ was found to be higher in
1978 and 1979, a result which is probably related to the
fact that in these years the income distribution behaves
differently (see the caption of figure \ref{fig1}).
\begin{figure*}
\epsfysize=16cm
\begin{center} \rotatebox{-90}{\epsffile{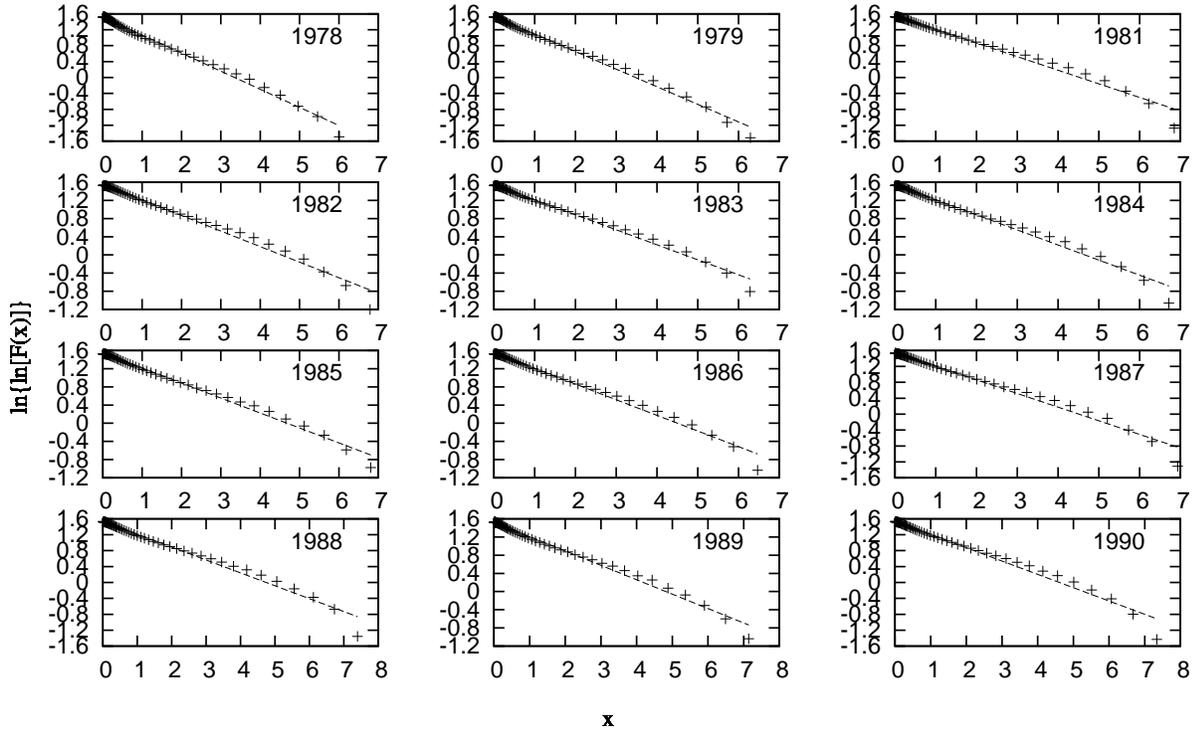}} \end{center}
\caption{Plots showing the fit of Gompertz curve to Brazil's
         individual income distribution data. The y-axis is the
	 double logarithm of the cumulative distribution, that
	 is, $\left\{ \ln \left[ \ln F \right] \right\}$, whereas
	 the x-axis is the normalized individual income $x$ up
	 to the value where $A \approx 1.5$. The dashed line is
	 the fitted straight line. Clearly the fit is good
	 up to very small values of $x$, a result which brings
	 support in favor of the Gompertz curve as a good model
	 for the income distribution of the economically less
	 favored individuals in the Brazilian population. Values
	 of the parameters resulting from the fit are presented
	 in table \ref{t-gomp}.\label{gomp1f}}
\end{figure*}
\begin{figure*}
\epsfysize=16cm
\begin{center} \rotatebox{-90}{\epsffile{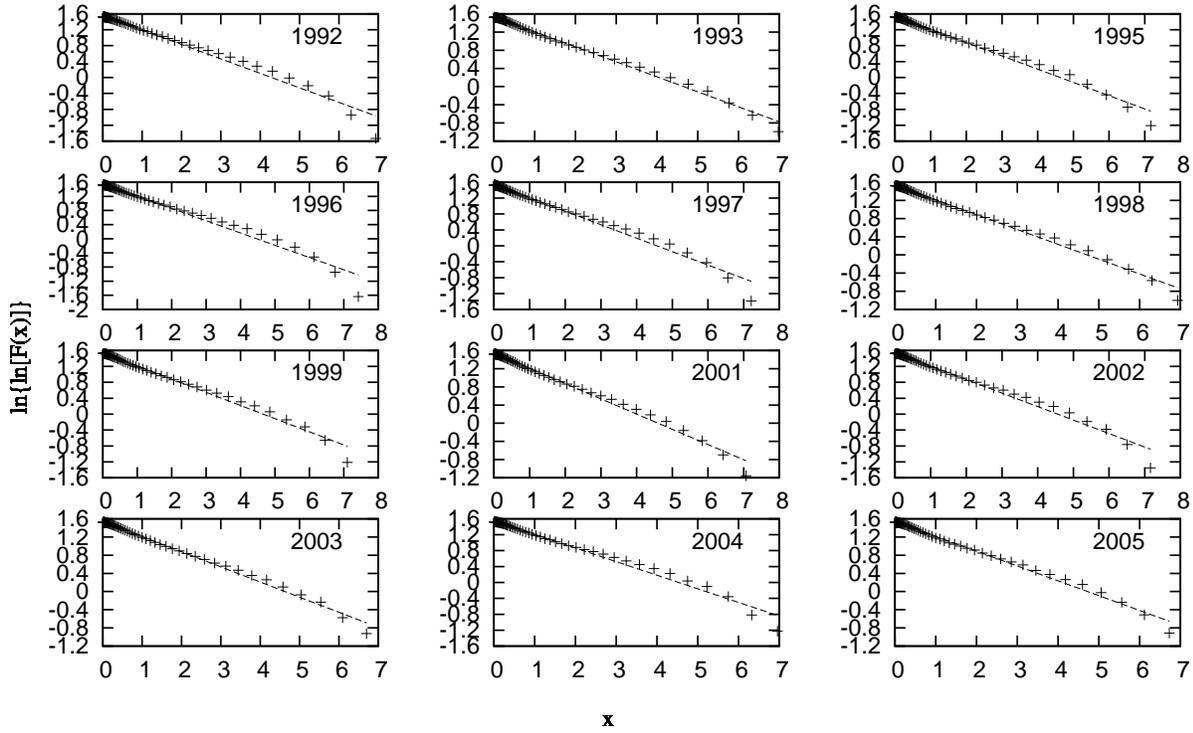}} \end{center}
\caption{Continuation of figure \ref{gomp1f} showing the fit of the
         Gompertz curve for data from 1992 to 2005.\label{gomp2}}
\end{figure*}
\begingroup
\begin{table*}[!htbp]
\caption{Results of fitting the Gompertz curve to Brazil's income
         distribution data from 1978 to 2005. The parameters were
	 obtained by least-square fitting and the respective errors
	 by means of one thousand bootstrap resamples with replacement
	 such that an average fitting and standard deviation
	 can be obtained for each sample in order to estimate the
	 uncertainties. Note the very good values of the correlation
	 coefficients of the fitting. Since the Gompertz parameters do
	 not change very much, we were able to reach an estimation valid
	 for the whole period of our analysis. That yields
	 $A=(1.54\pm0.03)$, $B=(0.39\pm0.08)$. Similarly, the end of
	 the Gompertz region is given by $x_{\mathrm{gmax}}=(7.4\pm0.8)$.
	 \label{t-gomp}}
\begin{center}
\begin{tabular}{cccccc}
\hline\noalign{\smallskip}
\textbf{year} & $\mathbf{A}$ & $\mathbf{B}$ &
$\mathbf{x_{{gmax}}}$ &
\textbf{correlation coeff.}
& \textbf{\% of individuals in Gompertz region} \\ 
\noalign{\smallskip}\hline\noalign{\smallskip}
1978 &$1.52\pm0.01$& $0.46\pm0.01$ &$6.606$&$0.997$&$98.9$\\
1979 &$1.54\pm0.01$& $0.44\pm0.01$ &$6.920$&$0.997$&$98.9$\\
1981 &$1.55\pm0.01$& $0.34\pm0.02$ &$7.533$&$0.992$&$98.9$\\
1982 &$1.55\pm0.01$& $0.34\pm0.02$ &$7.473$&$0.993$&$98.9$\\
1983 &$1.54\pm0.01$& $0.33\pm0.01$ &$6.910$&$0.996$&$98.7$\\
1984 &$1.55\pm0.01$& $0.33\pm0.01$ &$7.388$&$0.994$&$98.9$\\
1985 &$1.54\pm0.01$& $0.33\pm0.01$ &$7.490$&$0.996$&$98.9$\\
1986 &$1.55\pm0.01$& $0.34\pm0.01$ &$7.112$&$0.995$&$98.8$\\
1987 &$1.55\pm0.01$& $0.34\pm0.02$ &$7.626$&$0.992$&$98.9$\\
1988 &$1.54\pm0.01$& $0.32\pm0.02$ &$8.140$&$0.992$&$98.9$\\
1989 &$1.53\pm0.01$& $0.32\pm0.01$ &$7.856$&$0.995$&$98.8$\\
1990 &$1.54\pm0.01$& $0.34\pm0.02$ &$8.074$&$0.991$&$98.9$\\
1992 &$1.56\pm0.01$& $0.36\pm0.02$ &$7.635$&$0.989$&$99.0$\\
1993 &$1.54\pm0.01$& $0.33\pm0.01$ &$7.674$&$0.997$&$98.8$\\
1995 &$1.54\pm0.01$& $0.33\pm0.01$ &$7.887$&$0.995$&$98.9$\\
1996 &$1.55\pm0.01$& $0.35\pm0.02$ &$8.163$&$0.989$&$99.0$\\
1997 &$1.55\pm0.01$& $0.34\pm0.02$ &$7.935$&$0.992$&$99.0$\\
1998 &$1.54\pm0.01$& $0.33\pm0.01$ &$7.628$&$0.997$&$98.8$\\
1999 &$1.54\pm0.01$& $0.33\pm0.01$ &$7.811$&$0.994$&$98.9$\\
2001 &$1.54\pm0.01$& $0.34\pm0.01$ &$7.774$&$0.996$&$98.9$\\
2002 &$1.55\pm0.01$& $0.34\pm0.02$ &$7.878$&$0.993$&$99.0$\\
2003 &$1.54\pm0.01$& $0.33\pm0.01$ &$7.374$&$0.997$&$98.8$\\
2004 &$1.55\pm0.01$& $0.34\pm0.02$ &$7.653$&$0.993$&$98.9$\\
2005 &$1.54\pm0.01$& $0.33\pm0.01$ &$7.403$&$0.997$&$98.8$\\
\noalign{\smallskip}\hline
\end{tabular}
\end{center}
\end{table*}
\endgroup

\subsection{Parameters for the Pareto Law}

To fit the Pareto law we need to determine $x_{\mathrm{pmin}}$,
that is, the minimum value of $x$ that marks the start of the
Paretian part of the income distribution. In most years our
data clearly indicated that $x_{\mathrm{pmin}}$ ought to be
equal to $x_{\mathrm{gmax}}$. Nevertheless, due to the previously
discussed anomaly of the income distribution in 1978 and 1979,
the data for these years showed that $x_{\mathrm{pmin}} >
x_{\mathrm{gmax}}$. Inasmuch as from their definitions it is obvious
that $x_{\mathrm{gmax}} \le x_t \le x_{\mathrm{pmin}}$, for 1978
and 1979 the transition incomes between the Gompertzian and Paretian
regions and their uncertainties are evaluated as follows,
\be x_t=\frac{1}{2} \left( x_{\mathrm{pmin}} + x_{\mathrm{gmax}}
        \right), \; \;
    \delta x_t=\frac{1}{2}\left( x_{\mathrm{pmin}} -
     x_{\mathrm{gmax}} \right).
    \lb{xt}
\ee
Clearly if $x_{\mathrm{gmax}} = x_{\mathrm{pmin}}$, then
$x_t=x_{\mathrm{pmin}}$ and $\delta x_t=0$. These quantities were
then calculated in our dataset and the results are presented in
table \ref{t-pareto}.

The parameters $\alpha$ and $\beta$ were evaluated by two
different methodologies, least squares fitting and maximum
likelihood estimate. Details of both methods and comparison
of the results are described in what follows.

\subsubsection{Least Squares Fitting} 

This fitting method is not recommended when the data shows large
fluctuations, unless some binning process is employed such that
these fluctuations are severely reduced. As discussed in Section
\ref{data}, our data was treated that way and, therefore, we
believe that presenting the Pareto law parameters obtained by
least squares fitting (LSF) is useful, especially in order to
compare than with the other fitting method described below.

Figures \ref{paretof1} and \ref{pareto2} show the tail of the
complementary cumulative distribution where one can clearly
identify the power law decay in the data plots of all years.
These figures also show the straight line fitted by least
squares. Table \ref{t-pareto} presents the values of the
parameters found by LSF. Once can clearly notice that both
parameters of the Pareto law dot not remain as stable as the
parameters of the Gompertz curve during the time span of our
analysis. 

We must note that Cowell et al.\ \cite{cfl98} have previously
presented evidence for the Pareto law in the Brazilian individual
income distribution. Nonetheless, their study was restricted to
the shorter 1981-1990 period than the one considered here and,
even so, they only analyzed data for three years: 1981, 1985
and 1990. They assumed a log-normal distribution for the region
of lower income, but found out later that a Gaussian distribution
does not fit well the data. They also took the unusual step of
dividing the Pareto tail in two income range regions, one for
the rich and the other for the very rich, without presenting an
adequate justification for such a procedure, but reaching
conclusions about the ``increased inequality amongst the very
rich.'' This seems particularly odd if we bear in mind that the
income region of the very rich is exactly where we have the least
data and the statistical fluctuations are at their highest. As
stated above, here we present a study with a larger time span and
which includes all available data in the specified period,
totaling 24 yearly samples. We also advance the Gompertz curve
as a good descriptor for the lower individual income population
and found no evidence to support the claim made by Ref.\ \cite{cfl98}
of such two Paretian components. On the contrary, our data showed
very clearly a well defined and unique Pareto tail in all samples.

\subsubsection{Maximum Likelihood Estimation} 

This method is considered a better way of finding the Pareto index
because it deals well with the statistical fluctuations found in
the tails of income distributions. Here we shall closely follow
the approach proposed by Ref.\ \cite{n05} to derive the
likelihood of our dataset.

The constant $\beta$ is obtained as a result of the normalization
requirement (\ref{norm1}). As seen above, this normalization
is equivalent to the constraint equation (\ref{vinc}). Hence,
\be \beta = {x_{\sty t}}^{\displaystyle \alpha}
    e^{ \displaystyle e^{(A-Bx_{\sty t})} }.
  \lb{norm2}
\ee
This expression can be substituted into the probability density
(\ref{pareto1}), yielding,
\be p(x)= \alpha \; {x_{\sty t}}^{\alpha} e^{ \displaystyle
          e^{(A-Bx_{\sty t})} } \; x^{^{\scriptstyle
	  -(1+\alpha )}}.
    \lb{pareto3} 
\ee
The \textit{likelihood} of the data set is given by,
\be P(x|\alpha)=\prod_{j=1}^n p(x_j)=\prod_{j=1}^n 
    \alpha \; {x_{\sty t}}^{\alpha} e^{ \displaystyle
    e^{(A-Bx_{\sty t})} } \; {x_j}^{^{\scriptstyle -(1+\alpha )}}.
    \lb{prod}
\ee
We can calculate the most likely value of $\alpha$ by maximizing
the likelihood with respect to $\alpha$, which is the same as
maximizing the logarithm of the likelihood, denoted as
$\mathcal{L}$. Such calculation leads us to the following results,
\begin{eqnarray}
 \mathcal{L} & = & \ln P(x|\alpha) \nonumber \\
    & = & \sum_{j=1}^n \left[ \ln \alpha+\alpha \ln x_t
              + e^{(A-Bx_{\sty t})} - \left( 1+ \alpha \right) \ln
	                x_j \right] \nonumber \\
    & = & n \ln \alpha+ n \alpha \ln x_t
    + n e^{(A-Bx_{\sty t})} - \left( 1+ \alpha \right)
    \sum_{j=1}^n \ln x_j.
    \lb{lik1}
\end{eqnarray}
Setting $\partial \mathcal{L} / \partial \alpha =0$, we find,
\be \alpha=n {\left[ \sum_{j=1}^n \ln \left( \frac{x_j}{x_t}
           \right) \right] }^{-1}.
    \lb{lik}
\ee
Apart from a slight notation change, this result is equal to 
equation (B6) in Ref.\ \cite{n05}, despite the fact that this
work adopts a different normalization, as can be seen when
comparing equation (\ref{norm2}) above to equation (9) of
Ref.\ \cite{n05}. Therefore, this change in normalization
does not affect the estimation of the exponent of the Pareto
law obtained by the maximum likelihood estimator (MLE).

To find the expected error in the estimation of $\alpha$,
the width of the maximum of the likelihood as a function of
$\alpha$ should provide us with an estimate of $\delta \alpha$.
Taking the exponential of equation (\ref{lik1}) allows us
to find the likelihood as follows,
\be P(x|\alpha)= a e^{-b(1+ \alpha )} \alpha^n {x_t}^{n \alpha},
    \lb{lik2}
\ee
where
\be a= e^{\displaystyle \: n  e^{(A-Bx_{\sty t})}},
    \lb{a}
\ee
\be b= \sum_{j=1}^n \ln x_j \; .
   \lb{b}
\ee
Remembering that $\alpha>1$, the square root of the variance
in $\alpha$ will give us $\delta \alpha$. Therefore, we have
that,
\be \delta \alpha=\sqrt{\langle \alpha^2 \rangle - {\langle
                  \alpha \rangle }^2 },
    \lb{dalfa}
\ee
where
\be \langle \alpha \rangle = \frac{\displaystyle \int_1^\infty
    e^{-b(1+ \alpha )} \alpha^{(1+n)} {x_t}^{n \alpha} d \alpha}
    {\displaystyle \int_1^\infty e^{-b(1+ \alpha )} \alpha^n {x_t}^{n
    \alpha} d \alpha},
    \lb{med}
\ee
and
\be \langle \alpha^2 \rangle = \frac{\displaystyle \int_1^\infty
    e^{-b(1+ \alpha )} \alpha^{(2+n)} {x_t}^{n \alpha} d \alpha}
    {\displaystyle \int_1^\infty e^{-b(1+ \alpha )} \alpha^n {x_t}^{n
    \alpha} d \alpha}.
    \lb{med2}
\ee
Note that these two integrals can be solved numerically and that
both $x_j$ and $n$ in equations (\ref{lik}), (\ref{b}), (\ref{med})
and (\ref{med2}) refer only to the observed normalized income values
within the Pareto region, that is, $x_j \ge x_t$.

After calculating $\delta \alpha$, finding $\delta \beta$ becomes
just a matter of using standard error propagation techniques in
equation (\ref{norm2}). Table \ref{t-pareto} and figures
\ref{pareto_mv1} and \ref{pareto_mv2} present the results of
the Pareto law parameters obtained with the MLE.

\subsection{Percentage Populations and Percentage Share}

Once the Gompertzian and Paretian regions were established, we
were able to find the percentage of the population in each
component. The results shown in tables \ref{t-gomp} and \ref{t-pareto}
allowed us to determine that from 1978 to 2005 the income region
described by a Gompertz curve includes $(98.85\pm0.15)$\% of the
population of Brazil, whereas the Pareto region includes only
$(0.85\pm0.45)$\% of the Brazilian population. These results are
similar to the findings of Ref.\ \cite{dy01} for the USA, showing
that Brazil also has a two class income system where the overwhelming
majority of the population belongs to the lower income class.
\begin{figure*}
\epsfysize=16cm
\begin{center} \rotatebox{-90}{\epsffile{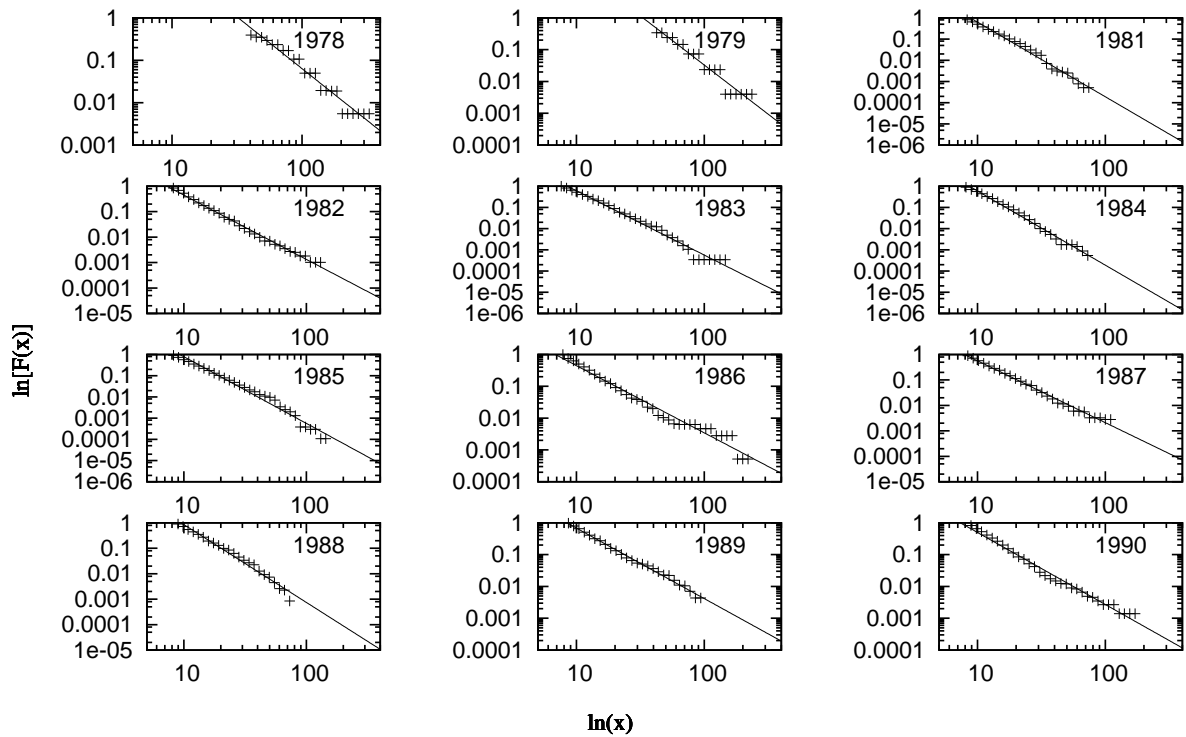}} \end{center}
\caption{Plots showing the least squares fitting (LSF) of the Pareto
         law to Brazil's income distribution data from 1978 to 1990.
         The full
         line is the fitted straight line. The Pareto power law is
         clearly visible in all plots and, differently from Ref.\
         \cite{cfl98}, we found no evidence of two Pareto tails. On the
         contrary, our plots show unique line fits. The tail fluctuations
         are visible, but they are not severe. Numerical values of the
         fitted parameters are shown in table \ref{t-pareto}.\label{paretof1}}
\end{figure*}
\begin{figure*}
\epsfysize=16cm
\begin{center} \rotatebox{-90}{\epsffile{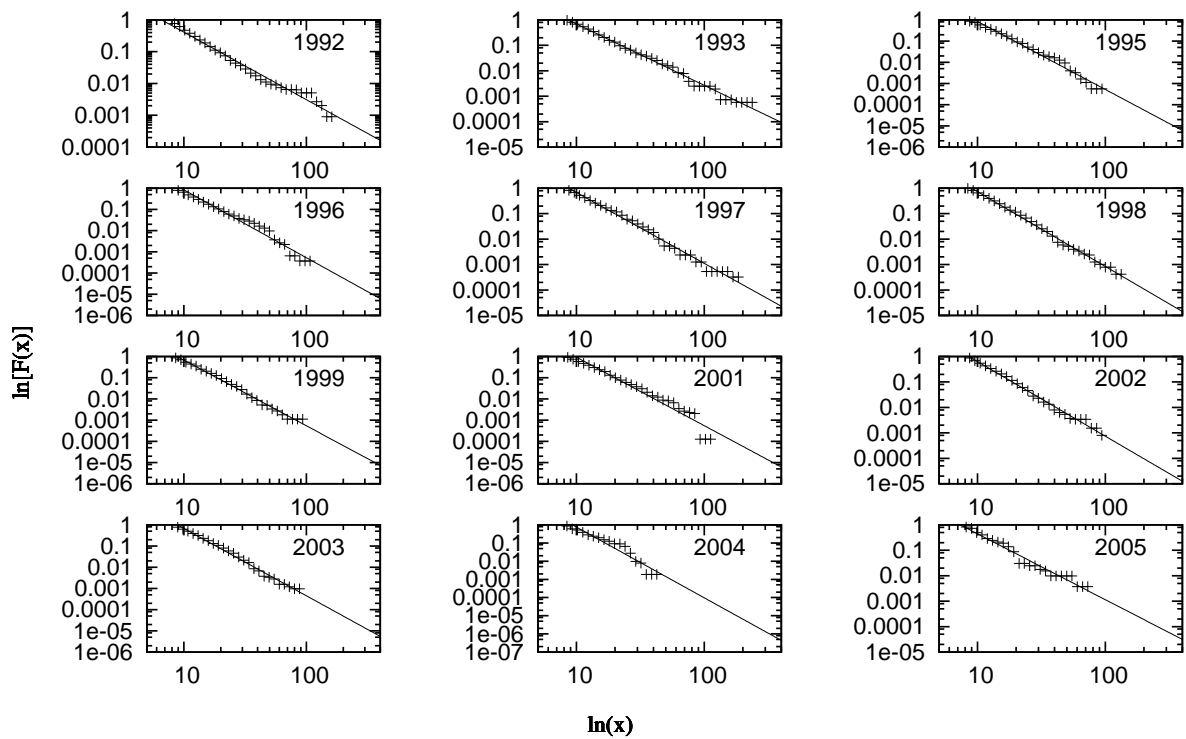}} \end{center}
\caption{Continuation of figure \ref{paretof1} showing the LSF of
         the Pareto power law for data from 1992 to 2005.\label{pareto2}}
\end{figure*}
\begingroup
\begin{table*}[!htbp]
\caption{Results of fitting the Pareto law to Brazil's income data. The
         transition values from Gompertz to Pareto regions are also shown.
         Due to the oddity of the data in 1978 and 1979 (see above),
	 there are indeed uncertainties of $x_t$ in these years. Results
	 of the parameters are presented by both methods, least squares
	 fitting (LSF) and maximum likelihood estimator (MLE). The
	 correlation coefficient obtained with LSF is included, as well
	 as the percentage of the population in the Paretian region.
	 $\delta \alpha$ and $\delta \beta$ in the LSF column were
	 calculated as in the Gompertzian region, that is, by bootstrap
	 replacement. The MLE method results in more stable values of
	 $\alpha$, whereas LSF results appear noisy.\label{t-pareto}}
\begin{center}
\begin{tabular}{cccccccccc}
\hline\noalign{\smallskip}
\textbf{year} & $\mathbf{x_{pmin}}$ & $\mathbf{x_t}$ 
 &\boldmath  $\mathbf{\alpha_{LSF}}$ &\boldmath  $\mathbf{\beta_{LSF}}$ 
 &\boldmath  $\mathbf{\alpha_{MLE}}$ &\boldmath  $\mathbf{\beta_{MLE}}$ &
 \textbf{c.c.\ LSF}
 & \textbf{\% in Pareto region}  \\
\noalign{\smallskip}\hline\noalign{\smallskip}
1978&40.0&$23.3\pm16.7$&$2.44\pm0.11$& $4767\pm4236$&$2.94\pm0.06$&$10639\pm22543$&$0.981$&$1.1-0.4$\\
1979&40.0&$23.5\pm16.5$&$3.09\pm0.17$&$49543\pm76479$&$3.09\pm0.09$&$17464\pm38230$&$0.973$&$1.1-0.4$\\
1981&7.533&$7.533$&$3.52\pm0.09$&$2071\pm781$&$2.84\pm0.11$&$444\pm100$&$0.993$&$1.1$\\
1982&7.473&$7.473$&$2.53\pm0.04$&$150.5\pm20.8$&$2.68\pm0.06$&$316\pm40$&$0.997$&$1.1$\\
1983&6.910&$6.910$&$3.03\pm0.08$&$655.6\pm186.6$&$2.64\pm0.05$&$263\pm26$&$0.992$&$1.3$\\
1984&7.388&$7.388$&$3.50\pm0.07$&$1870\pm451$&$2.84\pm0.11$&$441\pm97$&$0.996$&$1.1$\\
1985&7.490&$7.490$&$3.15\pm0.10$&$1112\pm382$&$2.66\pm0.05$&$312\pm34$&$0.990$&$1.1$\\
1986&7.112&$7.112$&$2.11\pm0.07$&$57.84\pm12.52$&$2.57\pm0.03$&$234\pm17$&$0.987$&$1.2$\\
1987&7.626&$7.626$&$2.43\pm0.06$&$153.6\pm26.7$&$2.72\pm0.07$&$360\pm55$&$0.996$&$1.1$\\
1988&8.140&$8.140$&$3.06\pm0.13$&$983.7\pm446.1$&$2.87\pm0.12$&$585\pm153$&$0.991$&$1.1$\\
1989&7.856&$7.856$&$2.22\pm0.04$&$112.5\pm11.4$&$2.78\pm0.09$&$445\pm80$&$0.997$&$1.2$\\
1990&8.074&$8.074$&$2.27\pm0.05$&$91.35\pm16.73$&$2.64\pm0.05$&$332\pm36$&$0.994$&$1.1$\\
1992&7.635&$7.635$&$2.12\pm0.06$&$54.52\pm10.36$&$2.64\pm0.05$&$288\pm31$&$0.990$&$1.0$\\
1993&7.674&$7.674$&$2.41\pm0.04$&$179.3\pm25.9$&$2.57\pm0.03$&$271\pm20$&$0.996$&$1.2$\\
1995&7.887&$7.887$&$3.21\pm0.10$&$1376\pm488$&$2.78\pm0.09$&$437\pm79$&$0.991$&$1.1$\\
1996&8.163&$8.163$&$3.20\pm0.12$&$1331\pm616$&$2.75\pm0.08$&$421\pm71$&$0.986$&$1.0$\\
1997&7.935&$7.935$&$2.79\pm0.05$&$419.3\pm79.7$&$2.62\pm0.04$&$310\pm32$&$0.995$&$1.0$\\
1998&7.628&$7.628$&$2.94\pm0.03$&$632.6\pm68.6$&$2.68\pm0.06$&$335\pm40$&$0.998$&$1.2$\\
1999&7.811&$7.811$&$3.10\pm0.07$&$867.5\pm205.5$&$2.78\pm0.09$&$429\pm77$&$0.996$&$1.1$\\
2001&7.774&$7.774$&$3.23\pm0.20$&$1588\pm1257$&$2.72\pm0.07$&$372\pm54$&$0.973$&$1.1$\\
2002&7.878&$7.878$&$2.93\pm0.05$&$539.8\pm89.1$&$2.78\pm0.09$&$426\pm79$&$0.996$&$1.0$\\
2003&7.374&$7.374$&$3.18\pm0.06$&$1017\pm227$&$2.78\pm0.09$&$387\pm68$&$0.996$&$1.2$\\
2004&7.653&$7.653$&$3.89\pm0.31$&$5793\pm19036$&$3.10\pm0.23$&$785\pm364$&$0.962$&$1.1$\\
2005&7.403&$7.403$&$2.59\pm0.09$&$172.7\pm50.8$&$2.84\pm0.11$&$441\pm97$&$0.984$&$1.2$\\
\noalign{\smallskip}\hline
\end{tabular}
\end{center}
\end{table*}
\endgroup
\begin{figure*}[tbh]
\epsfysize=16cm
\begin{center} \rotatebox{-90}{\epsffile{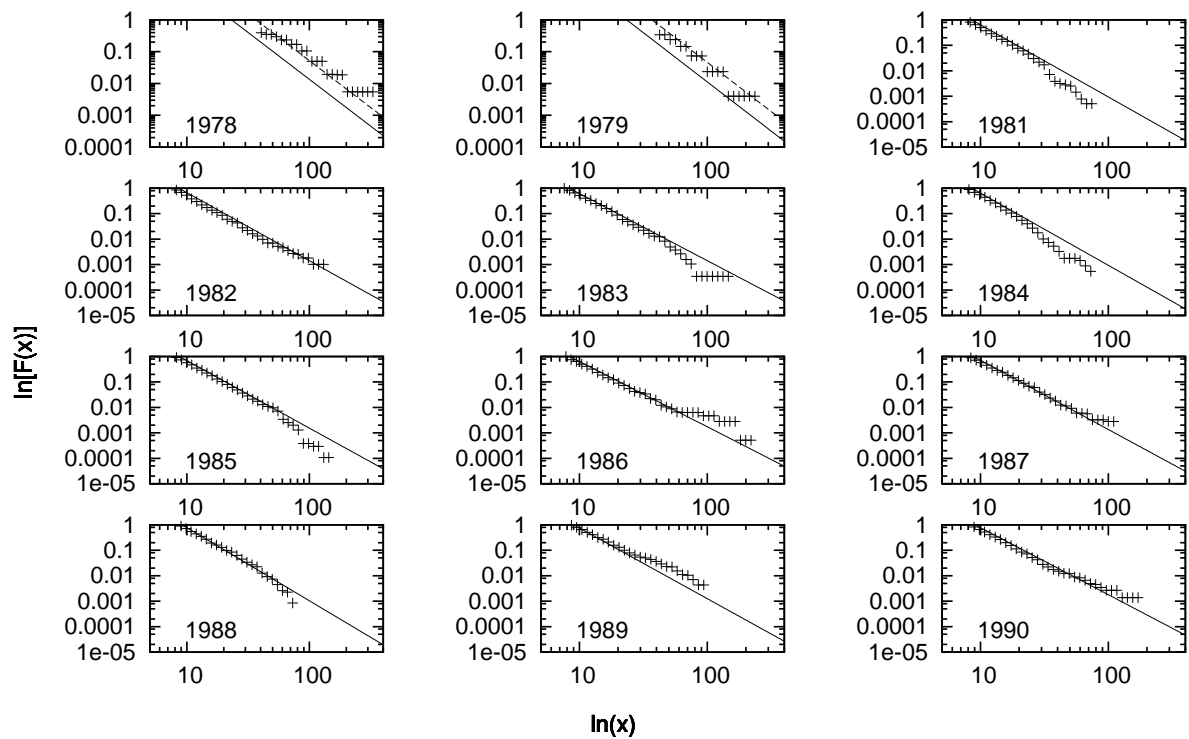}} \end{center}
\caption{These graphs present the tail of the complementary cumulative
         distribution fitted by a Pareto power law whose exponents
         were obtained with the maximum likelihood estimator (MLE). The
         dashed lines appearing in the plots of 1978 and 1979 represent
         the upper limits of $\delta \beta$ in these years, which,
         according to the results of table \ref{t-pareto}, are quite
         large.\label{pareto_mv1}}
\end{figure*}
\begin{figure*}[htb]
\epsfysize=15.4cm
\begin{center} \rotatebox{-90}{\epsffile{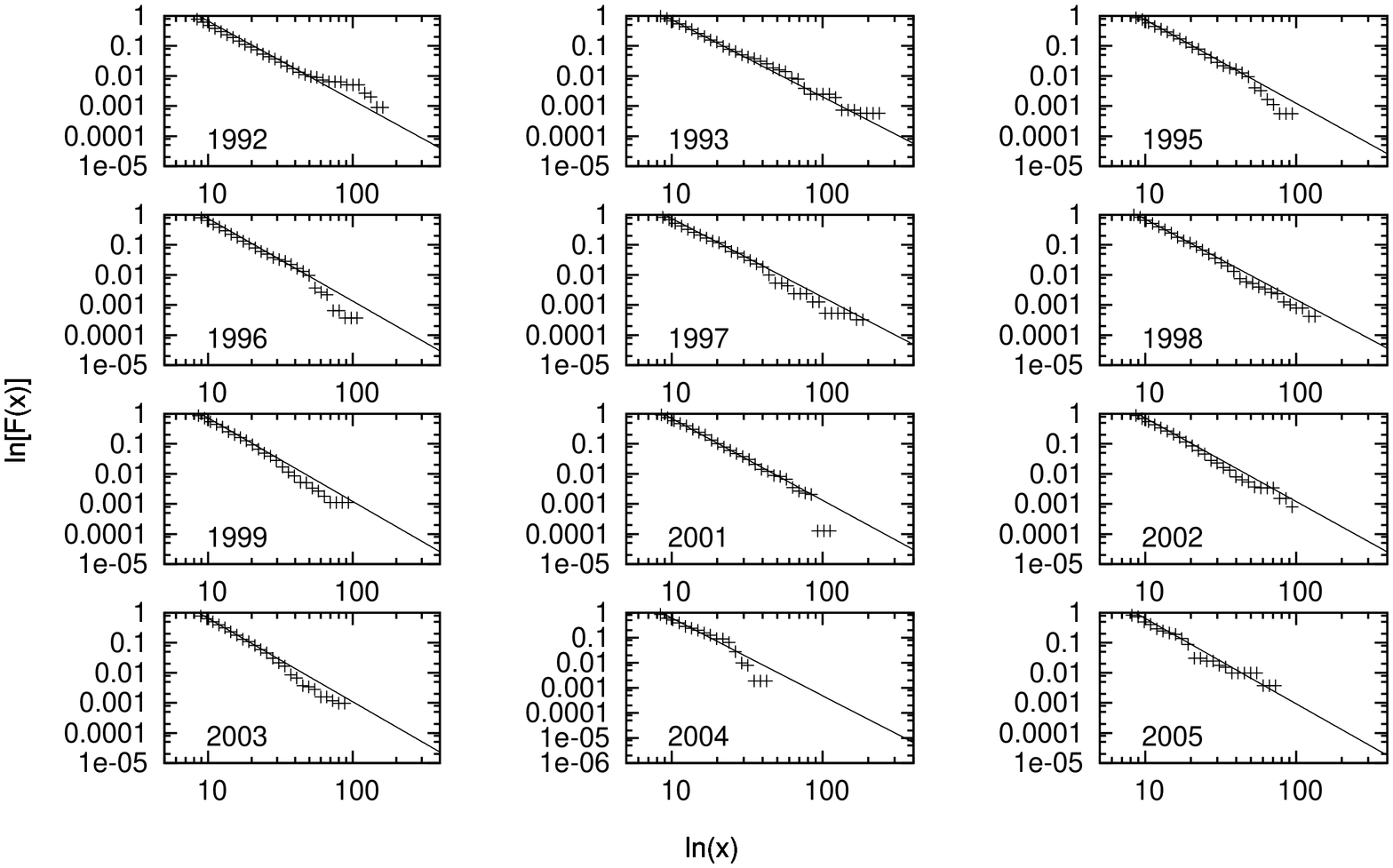}} \end{center}
\caption{Continuation of figure \ref{pareto_mv1} showing the Pareto
         power law fitted with the MLE from 1992 to
	 2005.\label{pareto_mv2}}
\end{figure*}

\begin{figure*}[htp]
\epsfysize=13cm
\begin{center}
\rotatebox{-90}{\epsffile{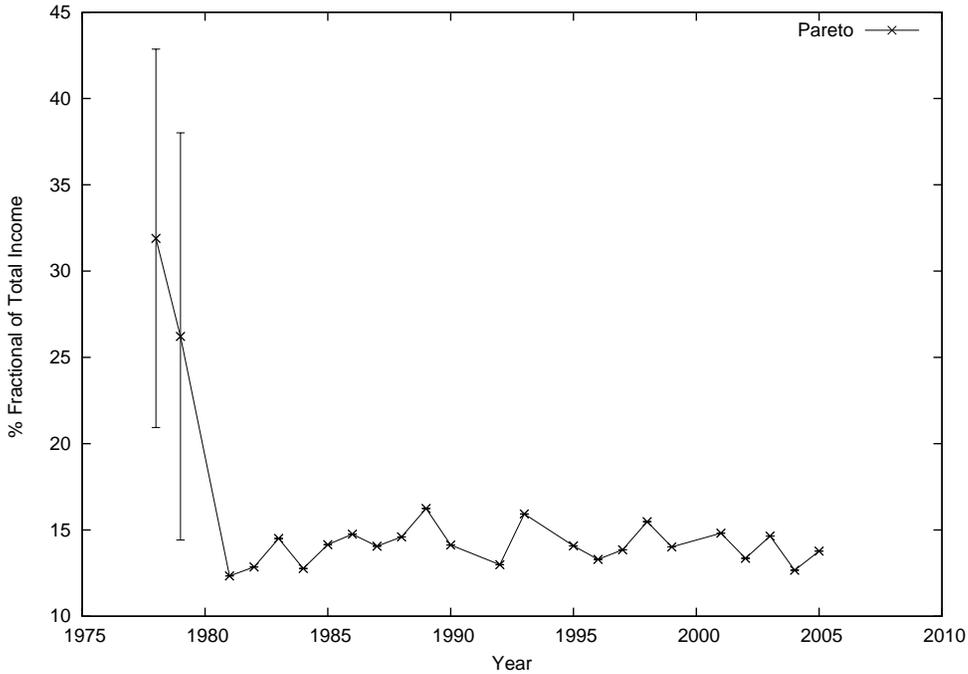}}
\end{center}
\caption{Plot showing the percentage share of the Pareto region
         relative to the total income. The error bars in 1978
	 and 1979 are due to the big uncertainty of the transitional
	 income value $x_t$ in these years, which led to huge
	 uncertainties in the income share of both the Gompertz and
	 Pareto regions. Even if we dismiss the portion due to these
         two years, it is apparent an irregular oscillatory pattern
	 with changing amplitude during the time span of our
	 analysis. The maximum and minimum inflexion points seem to
	 alternate roughly at every 2 years, whereas the period of
	 this oscillation occurs at approximately every 4 years.
	 If this oscillatory pattern is in fact a real feature of
	 the income distribution in Brazil, the next maximum of the
	 Paretian income share should occur in 2005-2007, while the
         next minimum should happen in 2007-2009. However, we must
         point out that \textit{this oscillatory pattern does not
         mean equilibrium}. There was economic growth for most of
	 the period shown here, albeit the growth rate was at
	 times fairly modest.\label{fig8}}
\end{figure*}

It is of interest to obtain the percentage share of each of the
two income components analyzed in this paper relative to the total
income. Table \ref{tab4a} presents these results together
with the Gini coefficients shown in figure \ref{fig5}. Due to the
same reasons discussed above the data analysis for 1978 and 1979 is
problematic because there is a large uncertainty in the transition
income $x_t$ between the Gompertzian and Paretian regions (see table
\ref{t-pareto}). Figure \ref{fig8} shows the percentage share of the
Pareto region and in this figure the uncertainties for 1978 and 1979
appear as large error bars for the first two points. If we dismiss
these two points, after a careful look at the irregular curve formed
by the variations of the Paretian percentage share we can see that
there is an oscillatory pattern, although with changing amplitudes,
whose periods can be set as roughly 4 years. The maximum and minimum
inflexion points seem to alternate at approximately every 2 years.

\begin{figure*}[htp]
\epsfysize=13cm
\begin{center}
\rotatebox{-90}{\epsffile{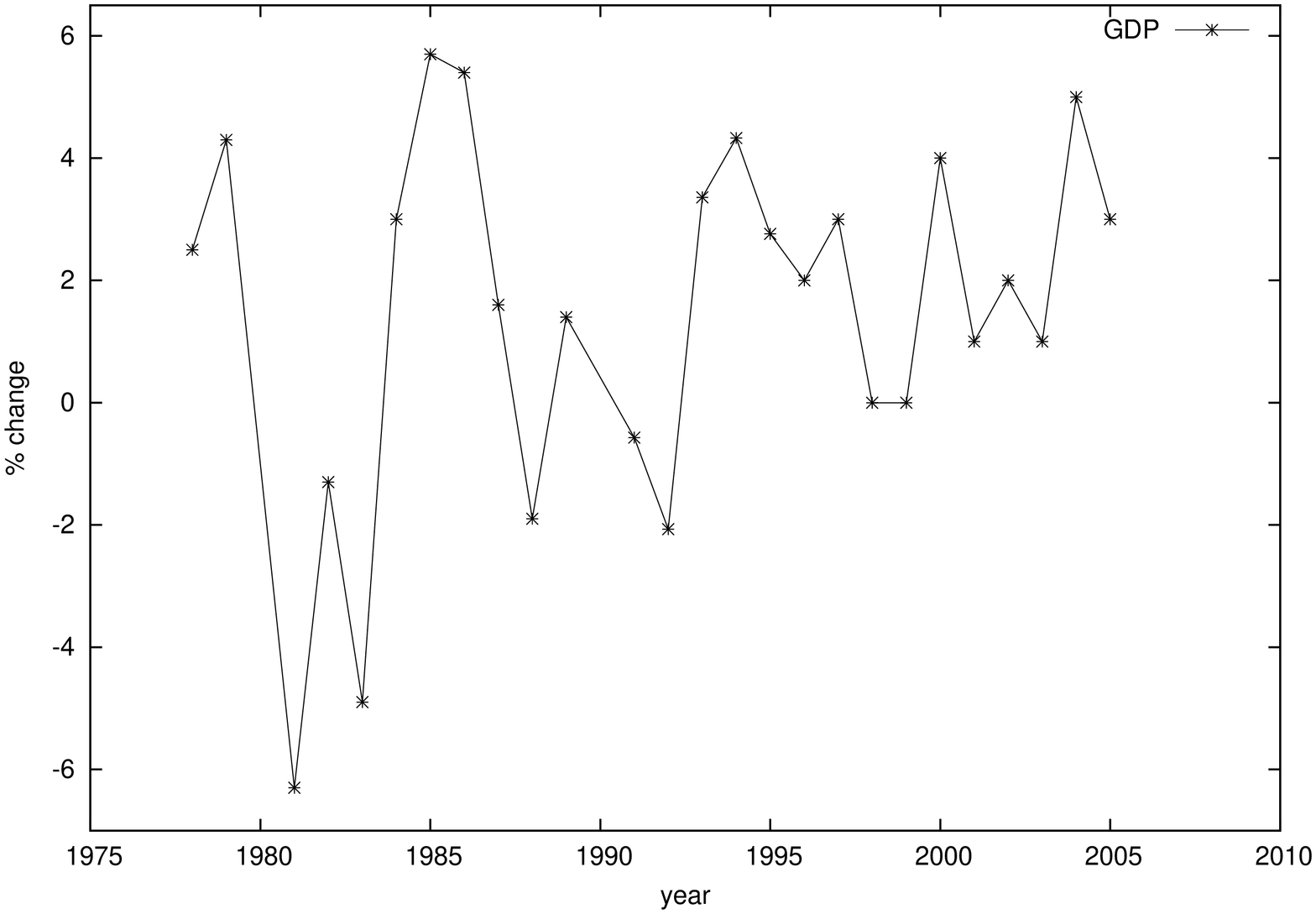}}
\end{center}
\caption{GDP growth rate in Brazil from 1978 to 2005. Although this
         graph also shows an approximate cycling pattern, the
	 oscillation shown here does not seem to correlate with the
	 cycle in the percentage share of the Paretian region
	 presented in figure \ref{fig8}.\label{gdp}}
\end{figure*}

It is interesting to know whether or not there is any possible
correlation of this approximate cycling pattern with any other
economic quantity. Figure \ref{gdp} presents a plot of the gross
domestic product (GDP) growth of Brazil in the same time period of
figure \ref{fig8} and, although we can also identify an approximate
cycling pattern in this graph, its oscillation does not seem to
correlate with the cycles in the percentage share of the Pareto
region.

As a final point, we should note that this approximate cycling
pattern in the Paretian share could be consistent with a purely
deterministic dynamical model based on the application of the
Lotka-Volterra equation to economic growth and cycle as advanced
long ago by Goodwin \cite{g67}. Although such a model predicts a
very regular oscillation of the percentage share of the lower
income class, this discrepancy with our data could perhaps be
remedied by the introduction of perturbation techniques. We
shall not pursue this issue further here \cite{mjr07}.
\begingroup
\begin{table*}[!htbp]
\caption{This table presents the percentage share relative to the
         total of each of the two components that characterize the
	 income distribution in Brazil. The Gini coefficients from
	 1978 to 2005 plotted in figure \ref{fig5} are also
	 presented. By definition these coefficients are obtained
	 as the area in between the two curves in figures \ref{fig3}
	 and \ref{fig4}. The area below each of the Lorenz curves was 
	 estimated numerically.\label{tab4a}}
\begin{center}
\begin{tabular}{cccc}
\hline\noalign{\smallskip}
\textbf{year} & \textbf{\% share of Gompertz region} &
\textbf{\% share of Pareto region} & \textbf{Gini coefficient}\\ 
\noalign{\smallskip}\hline\noalign{\smallskip}
1978 & 57.1 & 31.9 & 0.739  \\
1979 & 62.0 & 26.2 & 0.711  \\
1981 & 87.7 & 12.3 & 0.574  \\
1982 & 87.2 & 12.8 & 0.581  \\
1983 & 85.5 & 14.5 & 0.584  \\
1984 & 87.2 & 12.8 & 0.576  \\
1985 & 85.8 & 14.2 & 0.589  \\
1986 & 85.2 & 14.8 & 0.580  \\
1987 & 85.9 & 14.1 & 0.592  \\
1988 & 85.4 & 14.6 & 0.609  \\
1989 & 82.5 & 17.5 & 0.628  \\
1990 & 85.9 & 14.1 & 0.605  \\
1992 & 87.0 & 13.0 & 0.578  \\
1993 & 84.1 & 15.9 & 0.599  \\
1995 & 85.9 & 14.1 & 0.596  \\
1996 & 86.7 & 13.3 & 0.598  \\
1997 & 86.1 & 13.9 & 0.598  \\
1998 & 84.5 & 15.5 & 0.597  \\
1999 & 86.0 & 14.0 & 0.590  \\
2001 & 85.2 & 14.8 & 0.592  \\
2002 & 86.4 & 13.6 & 0.586  \\
2003 & 85.4 & 14.6 & 0.579  \\
2004 & 87.3 & 12.7 & 0.577  \\
2005 & 86.2 & 13.8 & 0.580  \\
\noalign{\smallskip}\hline
\end{tabular}
\end{center}
\end{table*}
\endgroup

\section{Conclusion}\lb{conclusion}

In this paper we have carried out an analysis of the personal income
distribution in Brazil from 1978 to 2005. We have made use of the
extensive household data surveys collected and made digitally available
by the Brazilian Institute for Geography and Statistics -- IBGE in order
to obtain 24 yearly samples of the complementary cumulative distribution
function $F(x)$ of individual income of Brazil in terms of the normalized
personal income $x$. We have concluded that this distribution function
is well described by two components. The first is a Gompertz curve of
the form $G\,(x)=\exp\,[\,\exp \, (A-Bx)]$, valid from $x=0$ up to
the transitional income $x_t$ and which includes $(98.85\pm0.15)$\%
of the population. The second component of the complementary cumulative
income distribution is a Pareto power law $P\,(x)= \beta\,x^{-\alpha}$,
valid from $x_t$ up. This includes the remaining $(0.85\pm0.45)$\% of
the population of Brazil. The positive parameters $A$, $B$, $\alpha$
and $\beta$ were all determined by a mixture of boundary conditions,
normalization and data fitting in all 24 yearly samples.
We also estimated uncertainties for these parameters. Lorenz curves
and Gini coefficients were also obtained, as well as the evolution of
the percentage share of both components relative to the total income.
The Paretian and Gompertzian shares show an approximate cycling pattern
with periods of about 4 years and maximum and minimum peaks alternating
at about every 2 years. These results show that the income distribution
pattern emerging from the present study allows us to
\textit{characterize Brazil as being formed by a well defined two class
system}.

The challenging questions posed by the results of this work concern 
the possible origins of the Gompertz curve. It seems quite reasonable
to suspect that the underlying dynamics of income distribution should
be intimately related to the dynamics of production and economic
growth in industrialized capitalist economies. Since economic growth
happens because production produces a net physical surplus, the
search for the origins of the Gompertz curve in income distribution
should perhaps focus in growth because this curve has been successfully
applied in models of population dynamics, particularly human
mortality from where it has originated \cite{sh05}, population
ecology \cite{kot01} and the growth of biomass \cite{ma06}. So,
the Gompertz curve may provide an important clue connecting income
distribution and economic growth as a result of net production
surplus. And although in these applications the power of the
first exponential of the Gompertz curve is negative whereas in
here it has a positive sign, such a difference may not be relevant
to the connection just mentioned. These remarks should also be
true for the \textit{logistic function}, which share with the
Gompertz curve the main feature of being S-shaped \cite{kot01,w32}
and also appears in economic models. From a physicists' standpoint,
it is well known that the dynamics of complex systems gives raise to
fractal power law patterns similar to the Pareto law. So, patterns
in economic growth, viewed perhaps as a complex dynamical system,
could be the root cause giving raise to the Gompertzian and Paretian
income distribution functions.

\begin{acknowledgement}
We would like to express our gratitude to Humberto Lopes, Jos\'e
Luiz Louzada, Vera Duarte Magalh\~aes and Cristiano de Almeida
Martins for their help with IBGE data. We are also grateful to two
referees for very useful comments which improved this paper.
\end{acknowledgement}

\end{document}